\documentclass[twocolumn, twocolappendix]{openjournal}

\usepackage{amsmath}

\usepackage{hyperref}
\hypersetup{colorlinks=true, linkcolor=blue, citecolor=blue, filecolor=blue, urlcolor=blue}

\newcommand{\ds}{\ensuremath{\Delta\Sigma}}
\newcommand{\gt}{\ensuremath{\gamma_{\mathrm{t}}}}
\newcommand{\rp}{\ensuremath{r_{\mathrm{p}}}}
\newcommand{\zl}{\ensuremath{z_{\mathrm{l}}}}
\newcommand{\zs}{\ensuremath{z_{\mathrm{s}}}}
\newcommand{\wsys}{\ensuremath{w_{\mathrm{sys}}}}
\newcommand{\wls}{\ensuremath{w_{\mathrm{ls}}}}
\newcommand{\sumls}{\ensuremath{\sum_{\mathrm{ls}}}}
\newcommand{\sigmacrit}{\ensuremath{\Sigma_{\mathrm{c}}}}
\newcommand{\mpch}{\ensuremath{h^{-1} \, \mathrm{Mpc}}}
\newcommand{\msunh}{\ensuremath{h^{-1} \, M_\odot}}

\begin{document}

\title{Systematic effects in galaxy--galaxy lensing with DESI}
\shorttitle{Systematic effects in galaxy--galaxy lensing with DESI}

\author{J.~U.~Lange,$^{1}$
C.~Blake,$^{2}$
C.~Saulder,$^{3,4}$
N.~Jeffrey,$^{5}$
J.~DeRose,$^{6}$
G.~Beltz-Mohrmann,$^{7}$
N.~Emas,$^{2}$
C.~Garcia-Quintero,$^{8}$
B.~Hadzhiyska,$^{6,9}$
S.~Heydenreich,$^{10}$
M.~Ishak,$^{8}$
S.~Joudaki,$^{11}$
E.~Jullo,$^{12}$
A.~Krolewski,$^{13,14,15}$
A.~Leauthaud,$^{10,16}$
L.~Medina-Varela,$^{8}$
A.~Porredon,$^{17,18,19}$
G.~Rossi,$^{20}$
R.~Ruggeri,$^{2,21}$
E.~Xhakaj,$^{10}$
S.~Yuan,$^{22}$
J.~Aguilar,$^{6}$
S.~Ahlen,$^{23}$
D.~Brooks,$^{5}$
T.~Claybaugh,$^{6}$
A.~de la Macorra,$^{24}$
P.~Doel,$^{5}$
K.~Fanning,$^{25,22}$
S.~Ferraro,$^{6,9}$
A.~Font-Ribera,$^{5,26}$
J.~E.~Forero-Romero,$^{27,28}$
E.~Gaztañaga,$^{29,11,30}$
S.~Gontcho A Gontcho,$^{6}$
S.~Juneau,$^{31}$
R.~Kehoe,$^{32}$
T.~Kisner,$^{6}$
A.~Kremin,$^{6}$
M.~Landriau,$^{6}$
M.~E.~Levi,$^{6}$
M.~Manera,$^{33,26}$
R.~Miquel,$^{34,26}$
J.~Moustakas,$^{35}$
E.~Mueller,$^{36}$
A.~D.~Myers,$^{37}$
J.~Nie,$^{38}$
G.~Niz,$^{39,40}$
N.~Palanque-Delabrouille,$^{41,6}$
C.~Poppett,$^{6,42,9}$
M.~Rezaie,$^{43}$
E.~Sanchez,$^{44}$
M.~Schubnell,$^{1,45}$
H.~Seo,$^{46}$
J.~Silber,$^{6}$
D.~Sprayberry,$^{31}$
G.~Tarl\'{e},$^{45}$
M.~Vargas-Maga\~na,$^{24}$
R.~H.~Wechsler,$^{25,47,22}$
Z.~Zhou,$^{38}$
and H.~Zou$^{38}$
}
\email{E-mail: julange.astro@pm.me}
\shortauthors{J.~U.~Lange et al.}

\begin{abstract}
    The Dark Energy Spectroscopic Instrument (DESI) survey will measure spectroscopic redshifts for millions of galaxies across roughly $14,000 \, \mathrm{deg}^2$ of the sky. Cross-correlating targets in the DESI survey with complementary imaging surveys allows us to measure and analyze shear distortions caused by gravitational lensing in unprecedented detail. In this work, we analyze a series of mock catalogs with ray-traced gravitational lensing and increasing sophistication to estimate systematic effects on galaxy--galaxy lensing estimators such as the tangential shear $\gt$ and the excess surface density $\ds$. We employ mock catalogs tailored to the specific imaging surveys overlapping with the DESI survey: the Dark Energy Survey (DES), the Hyper Suprime-Cam (HSC) survey, and the Kilo-Degree Survey (KiDS). Among others, we find that fiber incompleteness can have significant effects on galaxy--galaxy lensing estimators but can be corrected effectively by up-weighting DESI targets with fibers by the inverse of the fiber assignment probability. Similarly, we show that intrinsic alignment and lens magnification are expected to be statistically significant given the precision forecasted for the DESI year-1 data set. Our study informs several analysis choices for upcoming cross-correlation studies of DESI with DES, HSC, and KiDS.
\end{abstract}


\maketitle

\section{Introduction}

The large-scale structure of the Universe is a primary cosmological probe and particularly sensitive to cosmic structure growth, i.e., how the Universe evolved from its initial near isotropic state to forming highly non-linear structures such as stars, black holes, galaxies, clusters, and cosmic filaments. Large-scale galaxy surveys, such as the Sloan Digital Sky Survey \citep[SDSS;][]{Abazajian2009_ApJS_182_543} main galaxy sample, the Baryon Oscillation Spectroscopic Survey \citep[BOSS;][]{Dawson2013_AJ_145_10}, and the Extended Baryon Oscillation Spectroscopic Survey \citep[eBOSS;][]{Dawson2016_AJ_151_44}, probe the Universe's large-scale distribution of matter by measuring the positions and redshifts of thousands to billions of galaxies and other tracers such as quasars. Such surveys can be classified into spectroscopic and photometric surveys. Spectroscopic surveys measure detailed spectra for each object using, for example, highly multiplexed spectrographs. Contrary, photometric surveys only obtain coarse redshifts via broad-band photometry. By having detailed, high-precision redshifts, spectroscopic surveys can measure redshift-space distortions and infer cosmic structure growth via the Kaiser effect \citep{Kaiser1987_MNRAS_227_1}. Photometric surveys, on the other hand, can target many more galaxies and measure tiny shape distortions induced by weak gravitational lensing \citep[see, e.g.,][]{Bartelmann2001_PhR_340_291}, another probe of cosmic structure growth.

One of the leading spectroscopic surveys currently underway is conducted by the Dark Energy Spectroscopic Instrument \citep[DESI; ][]{Levi2013_arXiv_1308_0847, DESICollaboration2016_arXiv_1611_0036, DESICollaboration2016_arXiv_1611_0037, DESICollaboration2022_AJ_164_207, DESICollaboration2023_arXiv_2306_6308, DESICollaboration2024_AJ_167_62}. Similarly, ongoing state-of-the-art photometric weak lensing surveys include the Dark Energy Survey \citep[DES; ][]{Flaugher2015_AJ_150_150, Abbott2022_PhRvD_105_3520}, the Subaru Hyper Suprime-Cam Survey \citep[HSC; ][]{Aihara2018_PASJ_70_4, More2023_PhRvD_108_3520} and the Kilo-Degree Survey \citep[KiDS; ][]{Kuijken2015_MNRAS_454_3500, Asgari2021_AA_645_104}. A key cosmological parameter of interest is $S_8 = \sigma_8 \sqrt{\Omega_{\mathrm{m}} / 0.3}$, where $\sigma_8$ is the amplitude of matter fluctuations on a scale of $8 \, \mpch$ and $\Omega_{\mathrm{m}}$ is the matter density. Interestingly, several studies focusing on the large-scale structure of the low-redshift Universe have found tensions with the cosmic microwave background analysis \citep{PlanckCollaboration2020_AA_641_6}, the so-called $S_8$-tension \citep[see][and references therein]{Abdalla2022_JHEAp_34_49}. While the above-mentioned large-scale structure surveys are expected to produce leading constraints on cosmological parameters individually, a maximally informative cosmological analysis will need to analyze these surveys jointly.

One way to combine spectroscopic and photometric galaxy surveys is to measure the gravitational galaxy--galaxy lensing effect \citep{Brainerd1996_ApJ_466_623} around spectroscopically observed galaxies. In this case, we measure the mean shear distortions around spectroscopic galaxies, a measure of the mean mass distribution. This measurement is sensitive to $S_8$ as well as $b_{\mathrm{g}}$, the galaxy bias of spectroscopic galaxies. Combining these measurements with the clustering amplitudes of the same spectroscopic galaxies, which depend on $S_8$ and $b_{\mathrm{g}}^2$, one can solve the degeneracy between cosmology and galaxy bias and constrain cosmological parameters. Additionally, the combination of galaxy--galaxy lensing and galaxy clustering in redshift-space is a unique probe of modification to general relativity on cosmological scales. Studies of the galaxy--galaxy lensing effect have been employed widely to constrain cosmological parameters, particularly using spectroscopic data from SDSS combined with imaging data from other weak lensing surveys \citep[see, e.g.,][]{Mandelbaum2013_MNRAS_432_1544, Cacciato2013_MNRAS_430_767, Blake2020_AA_642_158, Heymans2021_AA_646_140, Lange2023_MNRAS_520_5373, Miyatake2023_PhRvD_108_3517}.

Gravitational lensing by foreground galaxies induces shear distortions in the apparent images of photometric galaxies behind it, giving rise to a non-zero mean tangential shear $\gt$. While this gravitational effect is most often the primary contributor to the apparent shape distortions around foreground spectroscopic galaxies, various systematic effects can also influence galaxy--galaxy lensing measurements. These effects include shear biases, magnification effects, incompleteness in the spectroscopic samples, photometric redshift dilution, and intrinsic alignment (IA). IA describes the intrinsic correlations between the shapes of nearby galaxies. As such, IA can mimic gravitational lensing and is one of the most significant sources of systematic errors in weak lensing. It has been studied extensively, particularly in the context of photometric surveys measuring the so-called cosmic shear, i.e., shape correlations of nearby galaxies. For an overview of IA see \cite{Lamman2024_OJAp_7_14} and for comprehensive reviews see \citep{Joachimi2015_SSRv_193_1, Kiessling2015_SSRv_193_67, Troxel2015_PhR_558_1}. We note that another important effect in the interpretation of gravitational lensing is baryonic feedback \citep[see, e.g.,][]{Leauthaud2017_MNRAS_467_3024, Lange2019_MNRAS_488_5771, Amodeo2021_PhRvD_103_3514, BeltzMohrmann2021_ApJ_921_112, Amon2023_MNRAS_518_477}. However, unlike the aforementioned effects, baryonic feedback alters the intrinsic matter field around the foreground galaxies. Thus, it is typically addressed by including baryonic physics in the modeling framework and will not be explored in the present paper which is concerned with potential biases in the measurement of lensing.

Systematic effects in galaxy--galaxy lensing measurements have been investigated in the past using theoretical calculations, simulations and observational data \citep[see, e.g.,][]{Mandelbaum2005_MNRAS_361_1287, Mandelbaum2013_MNRAS_432_1544, Miyatake2015_ApJ_806_1, Simet2015_MNRAS_449_1259, Leauthaud2017_MNRAS_467_3024, Speagle2019_MNRAS_490_5658, Unruh2020_AA_638_96, Prat2022_PhRvD_105_3528, Makiya2022_JCAP_03_008, Yao2023_AA_673_111, Amon2023_MNRAS_518_477}. However, given the high-precision measurements expected via a joint analysis of DESI, DES, HSC, and KiDS, further investigation is warranted. Our work extends previous studies by utilizing sophisticated, high-resolution simulations, the Buzzard mocks \citep{DeRose2019_arXiv_1901_2401}, that realistically simulate the above systematic effects. Additionally, many of the systematic effects discussed above depend on the data sets in question, such as the precision of photometric redshifts for the weak lensing survey or the nature and magnitude of incompleteness for spectroscopic samples. We are thus particularly interested in quantifying the importance of these systematic effects for combined studies with DESI, especially in the context of the upcoming DESI galaxy--galaxy lensing analysis. In addition to estimating the magnitude of these systematic effects, we will also investigate possible mitigation strategies. Given the statistical constraining power of cross-correlation studies of DESI with DES, HSC, and KiDS \citep{Yuan2024_arXiv_2403_0915}, we expect that the systematic error will need to be at the level of, at most, a few percent for cosmological studies to be statistically limited \citep{Yao2023_MNRAS_524_6071}. We also refer the reader to a companion study, \citet{Yuan2024_arXiv_2403_0915}, that focuses on another systematic, redshift evolution effects, in combined studies of DESI clustering and gravitational lensing. Finally, we note that detailed image simulations needed for characterizing shear biases and the effects of blending are beyond the scope of this work. Instead, those effects have been studied by the DES, HSC, and KiDS collaborations in their respective studies \citep{Mandelbaum2018_MNRAS_481_3170, Giblin2021_AA_645_105, MacCrann2022_MNRAS_509_3371}.

This paper is organized as follows. In section \ref{sec:estimators}, we outline the essential theoretical foundations of galaxy--galaxy lensing as well as the estimators we will use. Afterward, we give a brief overview of our simulated mock galaxy catalogs in section \ref{sec:mocks}. In section \ref{sec:results}, we show and discuss estimates of the different systematic effects in galaxy--galaxy lensing. We discuss our findings in the context of upcoming studies of DESI data cross-correlated with weak lensing surveys in section \ref{sec:discussion}. Finally, we present our conclusions in \ref{sec:conclusion}.

\section{Galaxy-galaxy lensing estimators}
\label{sec:estimators}

The deflection of light by the gravitational potential of a ``lens'' galaxy causes distortions in the apparent positions and shapes of background ``source'' galaxies. One way to probe this effect is by analyzing the apparent ellipticities of galaxies. In this section, we outline fundamental formalism and estimators used in galaxy--galaxy lensing.

\subsection{Theoretical background}
\label{sec:background}

Assuming circular symmetry\footnote{Even though individual mass distributions are not perfectly spherically symmetric, in the weak lensing regime, we stack a large enough number of lenses such that the mass distribution averaged over all lenses appears symmetric.} for the comoving lens surface mass density $\Sigma$, the mean reduced tangential shear $g_{\mathrm{t}}$ for a source separated by an angle $\theta$ from the lens is
\begin{equation}
    g_{\mathrm{t}}(\theta) = \frac{\gt(\theta)}{1 - \kappa(\theta)}\, ,
\end{equation}
where
\begin{equation}
    \kappa(\theta) = \frac{\Sigma(\theta)}{\sigmacrit (\zl, \zs)}
\end{equation}
is the convergence, $\zl$ and $\zs$ are the redshifts at the lens and source, respectively, and 
\begin{equation}
    \gt (\theta) = \frac{\ds(\theta)}{\sigmacrit (\zl, \zs)}
    \label{eq:gamma}
\end{equation}
is the shear. In the above equation, $\ds$ is the so-called excess surface density,
\begin{equation}
    \ds(\theta) = \bar{\Sigma}(<\theta) - \Sigma(\theta),
\end{equation}
where $\bar{\Sigma}(<\theta)$ the mean surface density inside for all angles smaller than $\theta$. Finally, the so-called critical surface density is defined via
\begin{equation}
    \sigmacrit (\zl, \zs) = \frac{c^2}{4\pi G} \frac{1}{(1 + \zl)^2} \frac{D_{\mathrm{A}}(\zs)}{D_{\mathrm{A}}(\zs) D_{\mathrm{A}}(\zl, \zs)}
\end{equation}
for $\zs > \zl$, where $D_{\mathrm{A}}$ denotes the angular diameter distance. For sources in front of lenses, i.e. $\zs \leq \zl$, the critical surface density can be defined to be $+\infty$ such that no shear distortion is induced, irrespective of the mass distribution around the lens. The factor $(1 + \zl)^{-2}$ accounts for the fact that we are working with comoving densities. In the weak lensing regime, $\kappa \ll 1$, it is common to approximate $g_{\mathrm{t}} \approx \gt$, the so-called reduced shear approximation.

\subsection{Tangential shear}

Individual galaxies, even in the absence of gravitational lensing, are typically not perfectly spherically symmetric, having a non-zero intrinsic ellipticity $e$. In the weak lensing regime, this intrinsic ellipticity of source galaxies is much larger than the gravitational shear $\gt$, an effect called shape noise. A commonly used galaxy--galaxy lensing estimator is the mean tangential gravitational shear $\gt$ as a function of angular separation $\theta$. Its estimator is
\begin{equation}
    \gt (\theta) = \overline{\mathcal{M}} (\theta) \left[ \gamma_{\mathrm{t, l}} (\theta) - \gamma_{\mathrm{t, r}} (\theta) \right] \, .
    \label{eq:gt_estimator}
\end{equation}
In the above equation, $\overline{\mathcal{M}}$ is a multiplicative shear calibration factor for ellipticity measurements. The estimator $\overline{\mathcal{M}}$ is specific to each of the three lensing surveys and will be discussed in more detail in section~\ref{subsec:shear_calibration}. Furthermore, $\gamma_{\mathrm{t, l}}$ and $\gamma_{\mathrm{t, r}}$ denote the ``raw'' tangential shear estimates around lenses and random points, respectively. They are defined via
\begin{equation}
    \gamma_{\mathrm{t, l}} (\theta) = \frac{\sumls \wsys \wls e_{\mathrm{s}}}{\sumls \wsys \wls} \, .
\end{equation}
for lenses and analogously for randoms where the sum goes over all suitable lens-source and random-source pairs in each angular bin. In the above equation, $\wsys$ is a systematic weight assigned to each lens galaxy to correct for selection biases such as fiber assignments. Similarly, $\wls = w_{\mathrm{s}}$ is the lens-source weight which, for $\gt$, reduces to the source weight $w_{\mathrm{s}}$ related to the uncertainty in the shape measurement, and $e_{\mathrm{t}}$ is the measured tangential ellipticity of the source galaxy with respect to the angle between lens and source. In the absence of shear systematics, $\gamma_{\mathrm{t, r}}$ should be statistically consistent with $0$. However, even in the absence of shear systematics, including this term can reduce cosmic variance uncertainties, especially at large angular separations \citep{Singh2017_MNRAS_471_3827}. Throughout this work, we measure $\gt$ in $15$ logarithmically-spaced bins going from $3$ to $300 \, \mathrm{arcmin}$.

\subsection{Excess surface density}

In addition to $\gt$, we will also study the $\ds$ estimator,
\begin{equation}
    \ds (\rp) = \overline{\mathcal{M}} (\rp) \left[ \ds_{\mathrm{l}} (\rp) - \ds_{\mathrm{r}} (\rp) \right] \, .
    \label{eq:ds_estimator}
\end{equation}
Here, $\rp = D_\mathrm{C} (\zl) \theta$ is the projected lens-source separation in comoving units. Analogous to $\gt$, $\ds_{\mathrm{l}}$ and $\ds_{\mathrm{r}}$ are the ``raw'' estimators for $\ds$ around lenses and randoms, respectively,
\begin{equation}
    \ds_{\mathrm{l}} (\rp) = \frac{\sumls \wsys \wls e_{\mathrm{t}} \Sigma_{\mathrm{ c, ls}}}{\sumls \wsys \wls} \, .
\end{equation}
Compared to the estimator for $\gt$, the $\ds$ estimator also includes the lens-source effective critical surface density $\Sigma_{\mathrm{ c, ls}}$, discussed in more detail in section~\ref{subsec:sigma_crit}. The lens-source weight $\wls$ is
\begin{equation}
    \wls = \frac{w_{\mathrm{s}}}{\Sigma_{\mathrm{ c, ls}}^\alpha} \, ,
    \label{eq:w_ls_ds}
\end{equation}
where we typically choose $\alpha=2$, which minimizes the impact of shape noise on the $\ds$-estimator \citep{Shirasaki2018_MNRAS_478_4277}. In this work, we measure $\ds$ in $15$ logarithmically-spaced bins going from $0.08$ to $80 \, \mpch$.

\subsection{Multiplicative shear calibration}
\label{subsec:shear_calibration}

$\overline{\mathcal{M}}$ is the multiplicative correction factor for the measured ellipticity of source galaxies. Different surveys utilize different algorithms and codes to estimate ellipticities and, therefore, $\overline{\mathcal{M}}$ has different estimators depending on the survey. Typically, lensing surveys also suffer from additive shear biases but those are not expected to impact cross-correlation functions like galaxy--galaxy lensing and we ignore them here. In the following, we will briefly describe the ellipticity measurements of different lensing data sets and the corresponding multiplicative correction factors.

\subsubsection{Dark Energy Survey}

Source ellipticities in the DES Year 3 (DES Y3) analysis are measured in the $riz$ bands using the {\sc NGMIX} \citep{Sheldon2015_ascl_soft_8008} code. The shear response matrix $\mathbf{R}$ defined as
\begin{equation}
    R_{ij} = \frac{\partial{\hat{e}}_i}{\partial e_j}
\end{equation}
is estimated using the {\sc METACALIBRATION} algorithm \citep{Huff2017_arXiv_1702_2600, Sheldon2017_ApJ_841_24} by artificially perturbing intrinsic ellipticities in the actual survey data. Further biases arise from selection biases \citep{Gatti2021_MNRAS_504_4312} and blending \citep{MacCrann2022_MNRAS_509_3371}. The latter two effects are small and so we ignore their impacts in the mock tests. For DES, modulo selection biases, and blending, the estimator for the shear calibration factor can be written as
\begin{equation}
    \mathcal{M}_{\mathrm{DES}} (\rp) = \frac{1}{\overline{R}_{\mathrm{t}} (\rp)} = \left( \frac{\sumls \wsys \wls R_{\mathrm{t}}}{\sumls \wsys \wls} \right)^{-1} \, ,
\end{equation}
where
\begin{equation}
    \begin{split}
        R_{\mathrm{t}} =& R_{11} \cos^2 (2 \phi) + R_{22} \sin^2 (2 \phi)\\&+ (R_{12} + R_{21}) \sin (2 \phi) \cos (2 \phi)
    \end{split}
\end{equation}
is the projection of the response matrix $R$ onto the vector connection lens and source and $\phi$ the polar angle of the source in the coordinate system of the source. We note that in contrast to HSC and KiDS, the shear response factors are estimated on a galaxy-by-galaxy basis. Particularly, individual estimates for $\mathbf{R}$ and $R_t$ can be very noisy. Since generally $\langle x \rangle^{-1} \neq \langle x^{-1} \rangle$, it is essential to compute ensemble-averaged estimates for $R_{\mathrm{t}}$ as correction factors for the ``raw'' $\gt$ and $\ds$ estimates instead of correcting each source by their individually estimated $R_{\mathrm{t}}$.

\subsubsection{Hyper Suprime-Cam}

The HSC lensing pipeline for the first-year analysis (HSC-Y1) is described in detail in \cite{Bosch2018_PASJ_70_5}. Galaxy ellipticities are measured using the {\sc HSM} algorithm \citep{Hirata2003_MNRAS_343_459}. Due to differences in the definition of galaxy ellipticity, the mean tangential ellipticity is expected to be $2 (1 - e_{\mathrm{rms}}^2) \gt$, where $e_{\mathrm{rms}}$ is the per-component intrinsic shape dispersion. The residual shear calibration, i.e., estimation of the multiplicative shear bias $m$ such that $\langle \hat{e} \rangle = 2 (1 - e_{\mathrm{rms}}^2) (1 + m) \langle e \rangle$, was performed in \cite{Mandelbaum2018_MNRAS_481_3170}. First, the authors estimated the ensemble shape dispersion $e_{\mathrm{rms}}$ for sources as a function of signal-to-noise ratio $S/N$ and resolution $R_2$. Then, the residual shear calibration is also estimated for sources binned by $S/N$ and $R_2$ and interpolated for each source.\footnote{\cite{Mandelbaum2018_MNRAS_481_3170} found that $e_{\mathrm{rms}}$ has a slight dependence on source redshift at fixed $S/N$ and resolution $R_2$. We ignore this correlation for the mock tests in this work.}

Given the above discussion, the shear calibration factor $\mathcal{M}$ for HSC is typically estimated as follows:
\begin{equation}
    \mathcal{M}_{\mathrm{HSC}} (\rp) = \frac{1}{2 \mathcal{R} (\rp) \left[ 1 + \overline{m} (\rp) \right]} \, ,
\end{equation}
where
\begin{equation}
    \mathcal{R} (\rp) = 1 - \frac{\sumls \wsys \wls e_{\mathrm{rms}}^2}{\sumls \wsys \wls}
\end{equation}
is the so-called shear responsivity and
\begin{equation}
    \overline{m} (\rp) = \frac{\sumls \wsys \wls m}{\sumls \wsys \wls} \,
    \label{eq:m}
\end{equation}
is the mean residual multiplicative shear bias. We note that the above estimator implicitly neglects correlations between $e_{\mathrm{rms}}$ and $m$. We will return to this point in section \ref{subsec:shear_calibration}. Finally, we note that our simulations will ignore the HSC selection bias that is primarily influenced by the cut on $R_2$. This effect is at the level of $\sim 1\%$ and would require image simulations to incorporate in the tests which goes beyond the scope of this work.

\subsubsection{Kilo-Degree Survey}

Galaxy ellipticities in the fourth data release of KiDS, the so-called KiDS-1000 data set, are inferred using a self-calibrating version of the {\sc lensfit} code \citep{Miller2007_MNRAS_382_315, Miller2013_MNRAS_429_2858, FenechConti2017_MNRAS_467_1627}. Residual multiplicative shear biases $m$ were determined as a function of tomographic redshift bin \citep{Hildebrandt2020_AA_633_69, Giblin2021_AA_645_105}, for ensembles of sources. Thus, for KiDS-1000, the estimator is simply
\begin{equation}
    \mathcal{M}_{\mathrm{KiDS}} (\rp) = \frac{1}{1 + \overline{m} (\rp)} \, ,
\end{equation}
where $\overline{m}$ is determined in the same way as for HSC, i.e. using equation (\ref{eq:m}).

\subsection{Critical surface density}
\label{subsec:sigma_crit}

To estimate $\ds$ as well as to interpret $\gt$, we need to estimate the effective critical surface mass density for lens-source pairs. However, unlike for DESI lenses, for sources, we typically lack precise, spectroscopic redshifts for all galaxies. Instead, we need to rely on imprecise photometric redshifts and determine the statistical relation between photometric and true redshifts \citep[see][for a review]{Newman2022_ARAA_60_363}. To do so, one can use sub-samples of source galaxies for which precise redshifts are available \citep[see, e.g.,][]{Buchs2019_MNRAS_489_820, Wright2020_AA_637_100}. Another method, the so-called clustering-redshift technique, is to study the spatial correlations of source galaxies with other targets with a known redshift distribution \citep[see, e.g.,][]{vandenBusch2020_AA_642_200, Gatti2022_MNRAS_510_1223}. For DES and KiDS, sources are divided into different tomographic bins, and the effective source redshift distribution $n(z)$ for each tomographic bin has been estimated. We note that the \textit{effective} $n(z)$, which is also needed for the cosmological interpretation of $\gt$, depends not only on the redshift distribution of the sources but also the redshift-dependent response to shear \citep[see, e.g.][for a detailed discussion]{MacCrann2022_MNRAS_509_3371}. For the HSC, an alternative approach is to use individual photometric redshifts. Thus, in this work, the estimators for $\Sigma_{\mathrm{c, ls}}$ for DES, KiDS, and HSC differ slightly.

We note that both methods using either tomographic bins or a calibration sample have statistical and systematic uncertainties related to noise or systematic bias in the calibration. In principle, ignoring some of the systematic effects such as the reduced shear, and assuming an unbiased calibration, both methods should lead to unbiased galaxy lensing estimates. In this work, we ignore uncertainties due to a noisy or potentially biased photometric-redshift calibration and assume that a perfectly calibrated $n(z)$ (for DES and KiDS) or calibration sample (for HSC) is available. We leave the investigation of the impact of the photometric redshift calibration to future work.

\subsubsection{Tomographic bins}

If effective source redshift distributions for each tomographic bin $t$ are known, the effective critical surface density becomes
\begin{equation}
    \Sigma_{\mathrm{c, ls}} (\zl, \zs) = \left[ \int\limits_0^\infty n_t(\zs) \sigmacrit^{-1} (\zl, \zs) \mathrm{d} \zs \right]^{-1} \, ,
    \label{eq:sigma_crit_eff}
\end{equation}
assuming the redshift distributions are normalized $\int n_t (\zs) \mathrm{d}\zs = 1$. In other words, for DES and KiDS, we calculate for each DESI lens of redshift $\zl$ a critical surface density that depends on the tomographic bin $t$ of the source galaxy, i.e., we ignore individual redshift estimates beyond tomographic bin associations. Finally, we note that the calibration of the effective source redshift distribution $n(z)$ has been performed for each tomographic bin as a whole. As a result, one cannot exclude or down-weight certain source galaxies within each tomographic bin as a function of their photometric redshift or other properties that may correlate with redshift.

\subsubsection{Calibration sample}

For HSC, we will work a calibration sample of source galaxies for which both wide-band photometric redshifts $\hat{z}_{\mathrm{s}}$ and high-quality redshifts $\zs$ are known. Assuming that the high-quality redshifts are an accurate representation of true redshifts, an unbiased estimate for the effective critical surface density is
\begin{equation}
    \begin{aligned}
        \Sigma_{\mathrm{c, ls}} &= \sigmacrit (\zl, \hat{z}_{\mathrm{s}}) \times \frac{\sum_{\mathrm{c}} w_{\mathrm{c}} \wls}{\sum_{\mathrm{c}} w_{\mathrm{c}} \wls \sigmacrit (\zl, \hat{z}_{\mathrm{s}}) / \sigmacrit (\zl, \zs)}\\
        &= \sigmacrit (\zl, \hat{z}_{\mathrm{s}}) \times f_{\mathrm{bias}} (\zl)\, .
    \end{aligned}
    \label{eq:f_bias}
\end{equation}
In the above equation, the sum goes over all sources in the calibration sample and $w_{\mathrm{c}}$ is a calibration weight (in addition to the usual source weight $w_{\mathrm{s}}$) that accounts for systematic biases in the calibration sample. A possible source of bias could be a non-representative calibration sample because of galaxy property-dependent targeting priorities and redshift success rates. Additionally, the calibration weights can account for a dependence on the shear response of each source. In other words, compared to DES and KiDS, for HSC we do use individual photometric redshift estimates $\hat{z}_{\mathrm{s}}$ but apply a statistical calibration factor $f_{\mathrm{bias}} (\zl)$ to the resulting estimate of the critical surface density. One advantage of this approach over calibrating the effective redshift distribution $n(z)$ of each tomographic bin is that one can apply further cuts on galaxies within each tomographic bin without biasing the lensing estimate.

\subsection{Lens magnification bias}
\label{subsec:lens_magnification}

Ideally, the angular distribution of lenses at a certain redshift $\zl$ would be independent of foreground structure at redshift $z_{\mathrm{f}} < \zl$. If that is the case, the shear field introduced by the foreground large-scale structure would only enter as noise into our $\gt$ and $\ds$ estimators. However, this assumption is not generally valid because the observed properties of lens galaxies are affected by the gravitational lensing of all objects between the observer and the lens. This can cause spatial correlations between foreground structure as well as its shear field and lens galaxy positions, an effect called lens magnification.

The lens magnification effect on the $\gt$ and $\ds$ estimator can be calculated analytically as follows. The foreground structure introduces a gravitational magnification field $\mu (\boldsymbol{\theta})$ of the lens galaxy plane. This has two effects on the lens galaxy population. First, this changes apparent lens galaxy position and alters the effective number density of lens galaxies by a factor $\mu^{-1}(\boldsymbol{\theta})$. Additionally, it changes the apparent brightness and size of individual objects. In turn, this can lead to individual objects falling in and out of the lens selection cuts, further altering the effective number density of lens galaxies.

Let us assume that a fraction $f_{\mathrm{l}} (\mu)$ of all galaxies passes the lens selection cuts. We define $\alpha_{\mathrm{l}}$ as the response of $f_{\mathrm{l}}$ to small changes in magnification,
\begin{equation}
  \alpha_{\mathrm{l}} = \left. \frac{\mathrm{d} \ln f_{\mathrm{l}}}{\mathrm{d} \mu} \right|_{\mu = 1} \, .
  \label{eq:alpha}
\end{equation}
In the weak lensing regime $|\mu - 1| \ll 1$, the change in the lens number density field can be written as
\begin{equation}
  \frac{n_{\mathrm{l}} (\boldsymbol{\theta}, \mu(\boldsymbol{\theta}))}{n_{\mathrm{l}} (\boldsymbol{\theta}, 1)} = 1 + (\alpha_{\mathrm{l}} - 1) \mu(\boldsymbol{\theta}) \, .
\end{equation}
\cite{Unruh2020_AA_638_96} show that this change in the lens density field introduces an additive bias in the mean tangential shear that can be estimated via
\begin{equation}
  \Delta \gt = 2 (\alpha_{\mathrm{l}} - 1) \gamma_{\mathrm{LSS}} (\theta, \zl, \zs) \, ,
\end{equation}
where
\begin{equation}
  \begin{split}
    \gamma_{\mathrm{LSS}} (\theta, \zl \zs) = &\frac{9 H_0^3 \Omega_{\mathrm{m}, 0}}{4 c^3} \int\limits_0^\infty \mathrm{d} \ell J_2(\ell \theta) \ell \int\limits_0^{\zl} \mathrm{d} z_{\mathrm{f}}\\
    &\frac{(1 + z_{\mathrm{f}})^2}{2 \pi} \frac{H_0}{H(z_{\mathrm{f}})} \frac{D_{\mathrm{A}} (z_{\mathrm{f}}, \zl) D_{\mathrm{A}} (z_{\mathrm{f}}, \zs)}{D_{\mathrm{A}} (\zl) D_{\mathrm{A}}(\zs)}\\
    &P_{\mathrm{m}} \left( \frac{\ell + 1/2}{(1 + z_{\mathrm{f}}) D_{\mathrm{A}} (z_{\mathrm{f}})}, z_{\mathrm{f}} \right) \, .
  \end{split}
\end{equation}
In the above equation, $J_2$ is the second order Bessel function, $P_{\mathrm{m}} (k, z)$ the non-linear matter power spectrum as a function of wavenumber and redshift, $H_0$ the Hubble parameter, $\Omega_{\mathrm{m}, 0}$ the matter density parameter, both at the present time.

\section{Mock catalogs}
\label{sec:mocks}

Our analysis is based on the Buzzard mock catalogs \citep{DeRose2019_arXiv_1901_2401} which, in turn, are based on a series of collision-less, dark matter-only simulations. The simulations have varying mass resolutions, ranging from $3.3 \times 10^{10} \msunh$ to $5.9 \times 10^{11} \msunh$, depending on the distance from the virtual observer. Galaxies are added to the dark matter-only simulations using the ``Adding Density Dependent GAlaxies to Lightcone Simulations'' technique \citep[ADDGALS; ][]{Wechsler2022_ApJ_931_145}. This approach gives virtual multi-wavelength galaxy catalogs with realistic number densities and clustering properties. Finally, gravitational lensing is simulated using the {\sc calclens} ray-tracing code \citep{Becker2013_PhDT_125} using $n_{\mathrm{side}} = 8192$, i.e., with an effective pixel size of $0.46$ arcmin. The lensing maps include the impact of lensing on shear as well as apparent positions and magnitudes of galaxy targets.

\subsection{DESI targets}

DESI targets various classes of extragalactic objects. Most relevant for galaxy--galaxy lensing studies are targets in the Bright Galaxy Survey (BGS) and luminous red galaxies (LRGs). Briefly, the BGS Bright sample includes galaxies above an apparent magnitude cut of $r < 19.5$ \citep{RuizMacias2020_RNAAS_4_187, Hahn2023_AJ_165_253}. BGS primarily targets galaxies in the redshift range $0.1 < z < 0.4$. In this work, we group BGS galaxies into three tomographic redshift bins: $0.1 \leq z < 0.2$, $0.2 \leq z < 0.3$, and $0.3 \leq z < 0.4$. Since BGS is a magnitude-limited sample, galaxy properties, and brightness can vary even within the narrow redshift bins. Thus, we reduce the sample by requiring a k-corrected rest-frame absolute magnitude $M_r > -19.5$, $-20.5$, and $-21$ for the three BGS redshift bins, respectively. The LRG sample targets fainter red galaxies in the redshift range of approximately $0.4 < z < 1.0$ \citep{Zhou2020_RNAAS_4_181, Zhou2023_AJ_165_58}. We divide the LRG sample into two tomographic bins, $0.4 \leq z < 0.6$ and $0.6 \leq z < 0.8$. We do not analyze LRGs at higher redshifts due to the lower LRG number density and scarcity of source galaxies at higher redshifts. We refer the reader to \citet{Yuan2024_arXiv_2403_0915} for a detailed discussion and motivation for these tomographic lens bins. In the mock catalogs, DESI targets are determined using the actual selection magnitude-dependent cuts used in observations \citep{DESICollaboration2024_AJ_167_62}. We also simulate the DESI redshift incompleteness by running the actual fiber assignment algorithm on the mock catalogs \citep{Silber2023_AJ_165_9}. Due to the DESI survey design, not all DESI targets will have redshifts which, if uncorrected for, will lead to systematic biases in the galaxy--galaxy lensing measurements compared to a complete survey.

Finally, we add random points to be used in the galaxy--galaxy lensing estimators. The random points follow the same redshift distribution $n(z)$ as the DESI targets, have random angular positions within the simulated mock survey and $3.3$ times the number density as the DESI tracers.

\subsection{Weak lensing targets}

Contrary to DESI targets, source galaxies from DES Y3, HSC-Y1, and KiDS-1000, have more complicated selection cuts and are added to the mock catalogs by matching the observed redshift and apparent magnitude distribution in real data. Similarly, photometric redshifts are assigned by matching the spectroscopic versus photometric distributions in observations. In particular, photometric redshifts are based purely on the true redshift of each source and do not explicitly depend on galaxy color. We refer the reader to DeRose et al. (in prep.) for a detailed discussion of this part of the mock-making procedure. The different tomographic bins for DES, HSC, and KiDS have been determined by the different weak lensing surveys. KiDS uses five tomographic bins whereas DES and HSC use four. The photometric redshift bin edges are (approximately) $[0.0, 0.358, 0.631, 0.872, 2.0]$, $[0.3, 0.6, 0.9, 1.2, 1.5]$ and $[0.1, 0.3, 0.5, 0.7, 0.9, 1.2]$ for DES, HSC and KiDS, respectively \citep{Hikage2019_PASJ_71_43, Myles2021_MNRAS_505_4249, Hildebrandt2021_AA_647_124}.

\subsection{Intrinsic alignment}

We model the IA of galaxies using a density-weighted Non-Linear Alignment model \citep[NLA;][]{Bridle2007_NJPh_9_444}. The NLA model is an evolution of the original Linear Alignment model \citep[LA;][]{Hirata2004_PhRvD_70_3526} which assumes that IA is linearly correlated with the gravitational potential.

As we only observe the alignment at positions of cosmologically-clustered source galaxies, the observed alignment is boosted (particularly at small physical scales); this corresponds to the clustering term in the tidal-torque alignment (TATT) model for intrinsic alignments~\citep{Blazek2019_PhRvD_100_3506}.

Following NLA, we can relate the convergence signal that would result from pure IA (with no lensing), $\kappa_{\textrm{IA}}$, linearly to the local density field via
\begin{equation}
    \kappa_{\textrm{IA}} (\hat{\boldsymbol{\mathrm{n}}}, z) = - A_{\textrm{IA}} C_1 \rho_{\textrm{crit}}  \frac{\Omega_{\mathrm{m}}}{D(z)}  \left( \frac{1+z}{1+z_0} \right)^{\eta_{\textrm{IA}}} \ \delta(\hat{\boldsymbol{\mathrm{n}}},  z)
\end{equation}
for some shell redshift $z$. We use the standard value of $C_1=5\times 10^{-14}M_{\odot}h^{-2}$Mpc$^2$. We choose a value of $A_{\textrm{IA}} = 0.5$ and assume no IA redshift evolution, so that $\eta_{\textrm{IA}}=0$.

To simulate an IA signal, we work with overdensity planes $\delta(\hat{\boldsymbol{\mathrm{n}}}, \chi)$, in thin redshift shells, which are represented as {\sc HEALPix} maps. These are obtained from maps of particle counts,
\begin{equation}
    \delta(\hat{\boldsymbol{\mathrm{n}}},  z) = n(\hat{\boldsymbol{\mathrm{n}}},  z)/ \langle n(\hat{\boldsymbol{\mathrm{n}}},  z) \rangle_{\mathrm{shell}}-1 \ \ ,
\end{equation}
where $n(z)$ is the number of particles in a shell with redshift $z$.

The $\kappa_{\textrm{IA}}$ maps are generated with the {\sc BornRaytrace} code\footnote{\url{https://github.com/NiallJeffrey/BornRaytrace}} using the simulated overdensity maps $\delta$. From these we generate shear maps $\gamma_{\textrm{IA}}$ that contain only an IA signal (i.e. no lensing), using the inverse Kaiser-Squire transformation on the sphere \citep[see][]{Kaiser1993_ApJ_404_441, Jeffrey2021_MNRAS_505_4626}, implemented within the {\sc BornRaytrace} code. That is, the spin-2 spherical harmonic coefficients of the shear, $\gamma_{\ell m}$ are related to the spin-0 spherical harmonic coefficients of the convergence, $\kappa_{\ell m}$, via
\begin{equation}
    \gamma_{\ell m} = -\sqrt{\frac{(\ell-1)(\ell+2)}{\ell(\ell+1)}} \kappa_{\ell m}.
    \label{eq:kappagamma}
\end{equation}
The resulting shear fields $\gamma_{\textrm{IA}}(\hat{\boldsymbol{\mathrm{n}}}, z)$ are used to assign the position-dependent IA amplitude to the ellipticity of galaxies in the mock catalog. 

The original density maps are generated at high resolution, i.e., {\sc nside}=4096 and with a thickness of $50 \, \mpch$, and shear maps were generated using a high multipole for the spherical harmonic transformations, $\ell$=8192; this ensures high accuracy at small angular scales (sub-arcmin). The above IA simulation and infusion were compared and validated using an independent pipeline using the spherical harmonics functionality from the {\sc Healpy} library to compute the projected tidal field and then directly apply the NLA model on each mass sheet by using said field components. The two infusions based on the NLA model were found to be in perfect agreement.

We note that different shape measurement algorithms may have different sensitivities to IA. This is because IA may depend on the galaxy radius at which the ellipticity is measured \citep{Singh2016_MNRAS_457_2301, MacMahonGeller2024_MNRAS_528_2980}. Thus, the different lensing surveys may show different IA sensitivities, even for the same galaxies, an effect we do not simulate here. Additionally, IA strength likely depends on galaxy properties such as color and luminosity \citep[see, e.g.][]{Yao2020_ApJ_904_135, Fortuna2021_MNRAS_501_2983} which may further contribute to different IA amplitudes in the different lensing surveys. Finally, it is worth mentioning that the NLA model used here may not describe IA on smaller scales, i.e., below $\sim 6 \mpch$, well \citep[see, e.g.,][]{Singh2015_MNRAS_450_2195}. More complex models are likely needed on these highly non-linear scales. Finally, while our choice of $A_{\textrm{IA}} = 0.5$ is consistent with recent cosmic shear studies \citep[see, e.g.,][]{Asgari2021_AA_645_104, Secco2022_PhRvD_105_3515}, the IA amplitude is poorly constrained with current data. We will return to these limitations of our IA model in section \ref{sec:discussion}.

\subsection{Gravitational signal}
\label{subsec:gravitational_signal}

In this work, we analyze a continuous $\sim 5000 \, \mathrm{deg}^2$ of mock sky catalogs, i.e., we do not match the actual footprints of the different surveys. The motivation is that in this work we aim to measure the impact of systematic effects with significantly higher precision than that of the galaxy--galaxy lensing measurements with DESI year-1 data. In particular, the simulated $5000 \, \mathrm{deg}^2$ overlap of DESI with DES Y3, HSC-Y1, and KiDS-1000 is significantly larger than the actual overlaps for (BGS, LRG) which are $(720, 850)$, $(140, 150)$, and $(450, 450)$ for DES, HSC and KiDS, respectively. We refer the reader to \citet{Yuan2024_arXiv_2403_0915}, where mock catalogs with matched footprints are used to estimate covariance properties of clustering and lensing measurements.

\begin{figure*}
    \centering
    {\Large\bf pure gravitational signal}
    \includegraphics{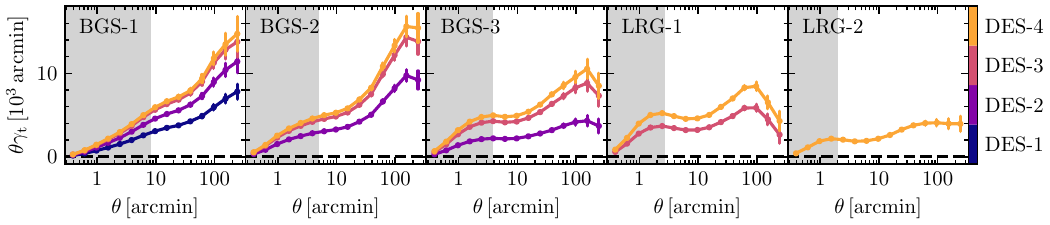}
    \includegraphics{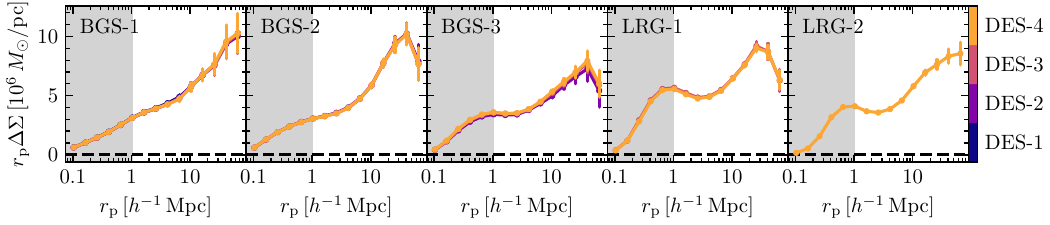}
    \caption{The mean tangential shear $\gt$ (top) and $\ds$ (bottom) around mock BGS and LRG samples cross-correlated with different tomographic source bins in DES. All systematic galaxy--galaxy lensing effects, including magnifications effects, fiber assignment, IA, and boost factors, are removed, i.e., the signal is a direct indicator of the mean gravitational potential around DESI targets. As expected, the $\ds$ amplitudes do not strongly depend on source tomographic bins. Results for HSC and KiDS are qualitatively and quantitatively similar. The shaded region denotes scales that we expect to be affected by resolution effects, as discussed in the text.}
    \label{fig:gravitational}
\end{figure*}

In Fig.~\ref{fig:gravitational}, we show idealized galaxy--galaxy measurements of DESI lens galaxies cross-correlated with DES lensing catalogs. Important systematic effects are turned off, including the impact of IA, magnification, and fiber incompleteness, such that we show the intrinsic gravitational signal we seek to recover. Error bars denote $1\sigma$ uncertainties derived from jackknife resampling of roughly equal-area patches of the simulated mock catalogs \citep{Shirasaki2017_MNRAS_470_3476}. As expected, $\gt$ depends on both the lens and source tomographic bin. Contrary, $\ds$ estimates depend primarily on the lens tomographic bin. However, there may be a small residual dependence of $\ds$ on the source tomographic bin. This is expected since the lens-source weighting in eq.~\eqref{eq:w_ls_ds} implies that the effective weighting by lens redshift $\zl$ depends on the effective source redshift $\zs$. As a result, cross-correlations with different source tomographic bins weigh lens galaxies differently depending on the redshift, giving rise to residual evolution effects \citep{Yuan2024_arXiv_2403_0915}.

\begin{figure}
    \includegraphics{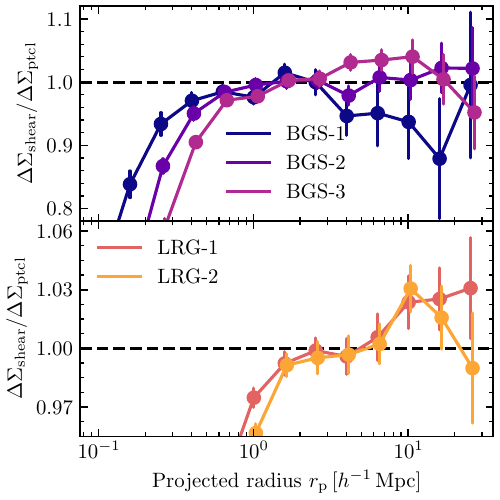}
    \caption{Ratio of the excess surface density $\ds$ estimated from the shear in the ray-tracing simulations to the $\ds$ estimate estimated directly from particles. Uncertainties on this ratio are derived from jackknife resampling.}
    \label{fig:ptcl_shear_ratio}
\end{figure}

Before proceeding, we want to estimate the resolution of our lensing which is affected by several factors. First, the limited resolution of the {\sc calclens} ray-tracing simulations will affect the minimum useful scale we can analyze. In Fig.~\ref{fig:ptcl_shear_ratio}, we compare the $\ds$ lensing amplitude estimated from the simulated ray-traced shear catalogs to $\ds$ estimated directly from downsampled particle catalogs. Differences between these two estimates should be primarily due to the limited resolution of the {\sc calclens} simulation. As expected, we find that these two estimates agree well on large scales but start to diverge once the separation becomes comparable to the effective resolution of the shear map, $\sim 0.46$ arcmin. In comoving coordinates, these effects always occur below a scale of around $1 \, \mpch$. The second effect is the limited resolution of the underlying dark matter simulations. \cite{DeRose2019_arXiv_1901_2401} argue that this is the primary effect limiting the resolutions of the Buzzard simulations and occurs at wavenumbers of $k \sim 2 \, h \, \mathrm{Mpc}^{-1}$ and higher. To convert this into a resolution limit for galaxy--galaxy lensing, we use the halo model used in \cite{Lange2019_MNRAS_488_5771}, truncate the power spectrum at $k \sim 2 \, h \, \mathrm{Mpc}^{-1}$ and calculate the resulting changes in $\Delta\Sigma$. This exercise indicates that resolution effects should start to become important below around $1 \, \mpch$. Finally, the {\sc addgals} algorithm is used to assign galaxies to the dark matter field in the simulations. Unlike other galaxy--halo connection models \citep[see][and references therein]{Wechsler2018_ARAA_56_435} such as Halo Occupation Distribution (HOD) or Subhalo Abundance Matching (SHAM), {\sc addgals} assigns most galaxies based on dark matter density instead of individual dark matter halos. However, this algorithm has been shown to produce realistic galaxy clustering down to $0.1 \, \mpch$ and is likely not a strong limiting factor in this study. Altogether, we estimate that scales below around $1 \, \mpch$ may show significant resolution effects. Therefore, we do not explore scales smaller than $1 \, \mpch$ in detail and warn the reader that higher-resolution simulations such as those described in \citep{Hadzhiyska2023_MNRAS_525_4367} may be needed to explore this regime.

\section{Galaxy-galaxy lensing systematic effects}
\label{sec:results}

In this section, we analyze various systematic effects in the galaxy--galaxy lensing estimators. In particular, we separate the impact of individual systematic effects on our estimate of the galaxy--galaxy lensing amplitude. To reduce statistical noise, we assume no shape noise. Additionally, the mock catalogs always cover the same continuous $\sim 5000 \, \mathrm{deg}^2$ on the sky, mitigating the impact of cosmic variance. In the following, uncertainties are derived by jackknife resampling of $100$ roughly equal-area patches on the sky. In this section, we focus on the relative impact of systematic effects. Thus, we only show results for combinations of lens and source bins that have a strong pure gravitational signal. That is the case when the majority of sources are behind the lenses. As a simple criterion, we demand that $\zs > \zl + 0.1$, where $\zs$ and $\zl$ are the median redshifts of each tomographic bin. For BGS, LRG, DES, HSC, and KiDS, the median redshifts in each tomographic bin are $[0.15, 0.25, 0.34]$, $[0.50, 0.72]$, $[0.30, 0.48, 0.71, 0.88]$, $[0.44, 0.75, 1.04, 1.30]$, and $[0.23, 0.39, 0.53, 0.75, 0.94]$, respectively.

\subsection{Shear calibration}

As described in section \ref{subsec:shear_calibration}, the shape algorithms employed by the different gravitational lensing surveys induce biases in the observed shear of source galaxies. Here, we verify that our galaxy--galaxy lensing estimators correctly account for such biases. To this end, we compare the $\ds$ estimates for sky catalogs with realistic levels of biases to ones without them, i.e. $m = 0$, $e_{\mathrm{rms}} = \sqrt{0.5}$ and $R_{ij} = \delta_{ij}$, depending on the survey, while turning off all other effects. We note that shear biases impact the effective redshift distribution of source galaxies since these are implicitly weighted by their response to the shear. Thus, even if we perfectly correct for shear biases, the presence of such biases would impact the measured $\gt$ by modifying the effective source redshift distribution. However, as long as the effects of the shear response are incorporated into the source redshift distribution, $\ds$ should remain effectively unchanged.

\begin{figure*}
    \centering
    {\large\bf residual shear bias}
    \includegraphics{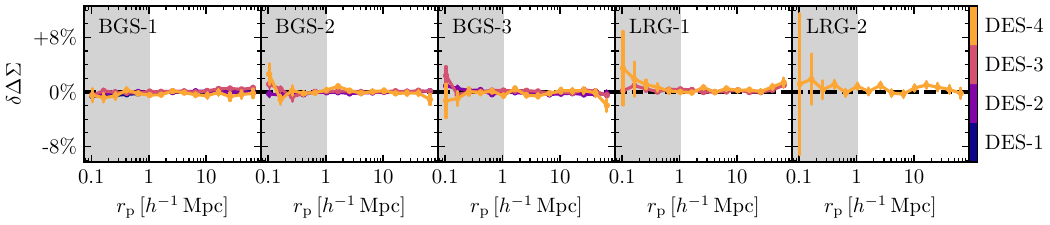}
    \includegraphics{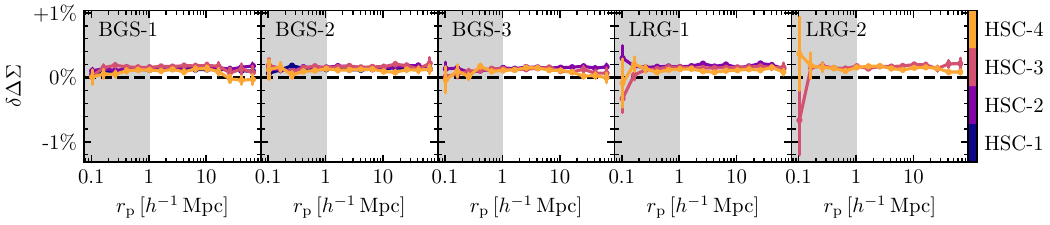}
    \caption{Residual biases in the $\ds$ estimator in the presence of shear biases in DES and HSC. To this end, we compare the measured galaxy--galaxy lensing amplitudes in mock catalogs with and without shear biases. We use galaxy--galaxy lensing estimators that are designed to correct for such biases. Residual biases for KiDS are exactly $0$ and thus are not shown here.}
    \label{fig:shear_bias}
\end{figure*}

In Fig.~\ref{fig:shear_bias}, we show DES and HSC estimates of residual biases in $\ds$ related to shear calibration after applying corrections. To this end, we calculate the relative difference in the measured $\ds$ amplitude for mocks with and without response biases. We find a slight residual bias for HSC of order $\sim 0.2 \%$ that is caused by the correlation between $e^{\mathrm{rms}}$ and $m$ that our estimator does not account for, i.e. $\langle (1 - e_{\mathrm{rms}}^2) (1 + m) \rangle \neq \langle 1 - e_{\mathrm{rms}}^2 \rangle \langle 1 + m \rangle$. Conversely, for DES and KiDS, we find no strong evidence for any residual bias in the estimators regarding correcting for shear response biases. Note that our estimates for KiDS are exactly $0$ since within each tomographic bin, the shear response bias is denoted by a single number, $m$, without scatter and can thus be corrected perfectly. This is likely an over-simplification since in real observations, $m$ will vary with galaxy properties.

In all three cases, the estimators do not exhibit any appreciable bias that would be relevant given the precision expected for DESI data. However, it is also worth noting that we implicitly assume that shear bias parameters for individual galaxies can be measured in an unbiased fashion. If, for example, the estimate of $\mathbf{R}$ returned from {\sc metacalibration} was biased, this would translate into additional biases on $\ds$ that are not captured in our analysis.

\subsection{Photometric redshift dilution}

To convert tangential ellipticities $e_{\mathrm{t}}$ into estimates of the excess surface density $\ds$, one needs to know the redshifts of the lenses and the sources. However, for source galaxies, we often only have noisy photometric redshifts for each source galaxy. For DES and KiDS, we measure $\ds$ by grouping source galaxies into tomographic bins and assigning each an effective critical surface density of the entire source galaxy population, as defined in eq.~\eqref{eq:sigma_crit_eff}. In other words, after binning into tomographic bins, the photometric redshift estimates for individual galaxies are no longer used. For HSC, we instead use the $f_{\mathrm{bias}}$-formalism whereby we use individual photometric redshift estimates for sources but later apply a correction factor based on a calibration sample, as described in eq.~\eqref{eq:f_bias}. Thus, for HSC, we can create mocks with perfect photometric redshifts, i.e., $\hat{z}_{\mathrm{s}} = \zs$, and establish a ground-truth $\ds$ amplitude measured from gravitational shear that should, in principle, be independent of the lensing survey and source tomographic bins.

We find that the galaxy--galaxy lensing amplitude is recovered without significant biases on large scales $\rp > 10 \mpch$. However, towards smaller scales, the lensing amplitude tends to be underestimated by a few percent. The reason is that while our estimators correctly account for the mean sky-averaged source redshift distribution, it does not take into account the increased fraction of sources physically associated with the lens, i.e., sources with $\zs \sim \zl$, close to lens galaxies. This effect, called the boost factor \citep[see, e.g.,][]{Kneib2003_ApJ_598_804, Sheldon2004_AJ_127_2544}, is discussed in the next section.

\subsection{Boost factors}
\label{subsec:boost}

\begin{figure*}
    \centering
    {\large\bf boost factor bias}
    \includegraphics{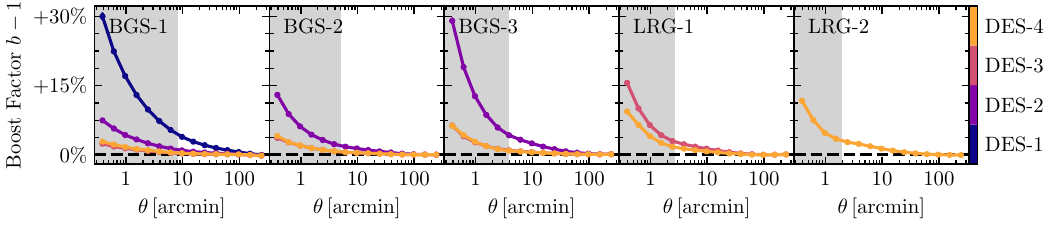}
    \includegraphics{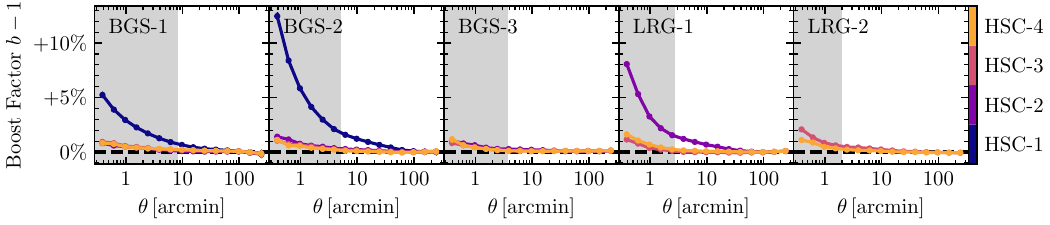}
    \caption{Boost factor estimates for $\gt$ in DES and HSC. The boost factors for $\ds$ are similar but slightly smaller since the $\ds$ estimator down-weights lens-source pairs with small redshift separations. Results for KiDS are similar to those from DES.}
    \label{fig:boost}
\end{figure*}

Due to galaxy clustering, there is an over-abundance of source galaxies at $\zs \sim \zl$ for small projected separations from lens galaxies. In turn, this excess of source galaxies with virtually no gravitationally induced shear from lens galaxies reduces the galaxy--galaxy lensing estimates. In the literature, this bias is often estimated via the so-called boost factor,
\begin{equation}
    b = \frac{\sum_{\mathrm{r}} \wsys}{\sum_{\mathrm{l}} \wsys} \frac{\sumls \wsys \wls}{\sum_{\mathrm{rs}} \wsys w_{\mathrm{rs}}} \, ,
\end{equation}
where $\sum_{\mathrm{l}}$ and $\sum_{\mathrm{r}}$ represent sums over all lens and random targets, respectively. The above ratio of weighted lens-source and random-source pair counts acts as an estimator of the lens-source clustering. In the case of physical lens-source clustering, we would expect $b > 1$ with $b \rightarrow 1$ in the limit $\rp \rightarrow \infty$. We would then correct for physical lens-source associations by adjusting the lensing amplitude around lenses (but not randoms), e.g., $\ds_{\mathrm{l}} \rightarrow b \ds_{\mathrm{l}}$. In Fig.~\ref{fig:boost}, we show estimates of the boost factor obtained from mocks with photometric redshifts but without the effects of magnification, fiber incompleteness, etc. As expected, we see that the effect is the largest on small scales and for lens-source bin pairs that are close together in redshift. Similarly, boost factors are stronger for DES compared to HSC and KiDS (not shown). This is largely due to DES photometric redshifts being based on three-band photometry whereas HSC and KiDS use more bands. This leads to more precise redshifts for the latter, minimizing the redshift overlap of lenses and sources.

In principle, we could correct our galaxy--galaxy lensing estimators using the boost factor. The problem is that other physical and observational effects can lead to lens-source clustering. For example, source magnification, the magnification of source galaxies by lens galaxies, can lead to an increase in the apparent number density of source galaxies close to lens galaxies. Such an increase would bias our estimator of $b$. Another effect that might bias the boost factor estimate is obscuration and blending by galaxies close to the lens \citep{Simet2015_MNRAS_449_1259}. Similarly, lensing quality cuts due to contamination of the background by the intracluster light associated with the foreground lens galaxies can further bias this estimate \citep{Leauthaud2017_MNRAS_467_3024, Everett2022_ApJS_258_15}. As we will show in section \ref{sec:discussion}, boost factors are unlikely to be significant for the DESI year-1 analysis as long as one analyses scales above a few $\mpch$ or only uses tomographic source bins that have little overlap with lens bins.

\subsection{Intrinsic alignments}

\begin{figure*}
    \centering
    {\large\bf intrinsic alignment bias}
    \includegraphics{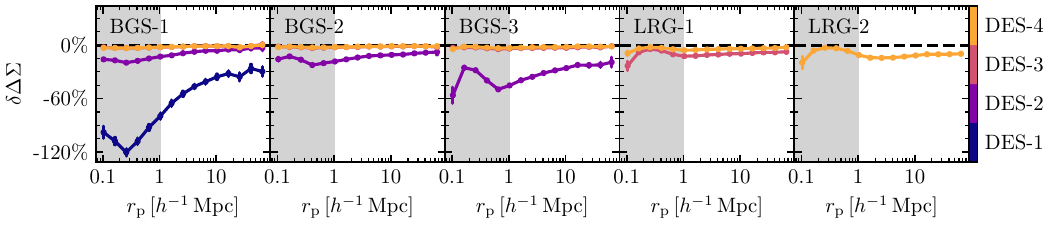}
    \includegraphics{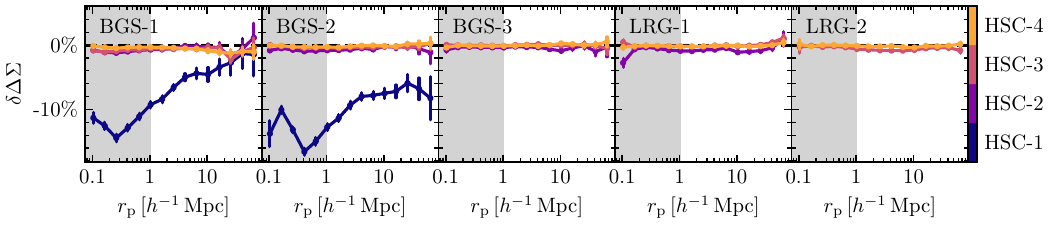}
    \includegraphics{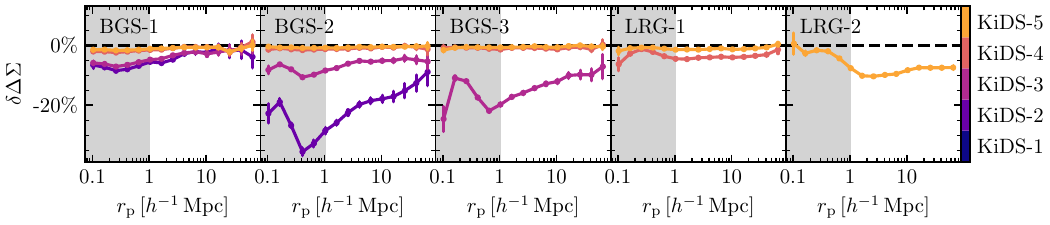}
    \caption{Relative impact of IA on the $\ds$ galaxy--galaxy lensing estimator. The impact on $\gt$ is similar but larger since the $\ds$ estimator down-weights lens-source pairs with small redshift separations.}
    \label{fig:intrinsic_alignment}
\end{figure*}

Galaxies are expected to have preferential alignment with their surrounding tidal field and large-scale structure. This IA effect causes non-zero shear correlations of nearby galaxies even in the absence of gravitational lensing. Traditionally, IA has been of particular relevance for estimates of cosmic shear. However, given the increased accuracy of the galaxy--galaxy measurements with DESI, IA contamination also becomes relevant for estimates of $\gt$ and $\ds$. The basic idea is that a non-negligible fraction of source galaxies will be at a similar redshift to the lens galaxies and may have a non-zero intrinsic tangential ellipticity with respect to the lens. Thus, this effect is particularly relevant if there is a large overlap between lenses and source in redshift.

In Fig.~\ref{fig:intrinsic_alignment}, we show the relative impact of IA on the $\ds$ galaxy--galaxy lensing estimator by comparing mocks with and without IA. We find that IA reduces the measured galaxy--galaxy lensing signal. Similar to the boost factor, the effect is larger for source tomographic bins with larger overlaps with the lens bins. Similarly, the effect is larger for DES than HSC and KiDS since the latter two have more precise redshift estimates. However, the effect of IA is much larger and, unlike boost factors, the effect of IA extends out to larger separations. Thus, it is crucial to model IA if one uses galaxy--galaxy lensing for cosmology. At the same time, our results indicate that the impact can be reduced down to $\sim 1\%$, at least for BGS, if more aggressive lens-source cuts are employed. Unfortunately, the effect on LRG measurements is stronger, especially with DES Y3 and KiDS-1000 lensing catalogs, and may be significant even for stringent lens-source cuts.

\subsection{Fiber assignment}

\begin{figure*}[h]
    \centering
    {\large\bf fiber assignment bias}
    \includegraphics{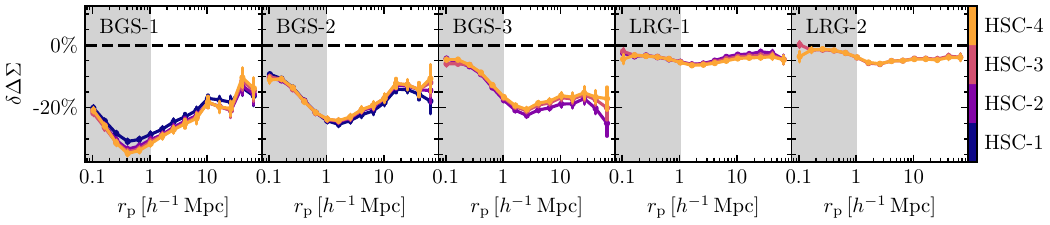}
    \caption{Relative impact of the fiber assignment on the measured galaxy--galaxy lensing amplitude with HSC. For this figure, no corrections for fiber incompleteness have been incorporated in the estimator. The estimator is biased since DESI will have relatively less completeness in overdense regions of the sky. Results for DES and KiDS are qualitatively and quantitatively very similar.}
    \label{fig:fiber_assignment_no_iip}
\end{figure*}

DESI will only measure spectroscopic redshifts for a subset of BGS and LRG targets. The limiting factor is the density of fibers on the DESI focal plane such that targets in overdense regions tend to be under-sampled. Due to the strong correlation of local density, large-scale bias, and halo mass, measuring the lensing amplitude around targets with spectroscopic redshifts can lead to strong biases. We test this effect by creating mocks that have the DESI fiber assignment algorithm applied to them. Here, we simulate the completed DESI survey with each portion of the sky receiving multiple passes, increasing the completeness of targets for which redshifts are obtained. The impact of fiber incompleteness would be even stronger for an incomplete DESI survey where certain parts of the sky are only covered by one pass, as is the case of the DESI year-1 data set. In Fig.~\ref{fig:fiber_assignment_no_iip}, we show the change in the measured lensing amplitude once only the subset of targets with spectroscopic redshifts are considered. We see that the lensing amplitude of BGS targets would be significantly under-estimated, up to $30\%$, owing to the high spectroscopic incompleteness of this target class. Contrary, the effect is smaller, at the level of $\sim 5\%$, for the spectroscopically more complete LRG sample.

\begin{figure*}[h]
    \centering
    {\large\bf residual fiber assignment bias}
    \includegraphics{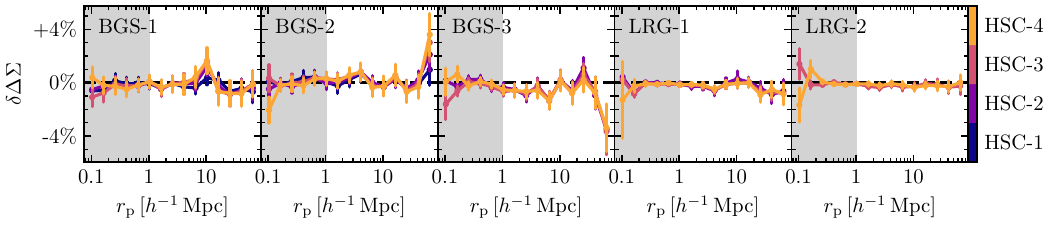}
    \caption{Same as Fig.~\protect\ref{fig:fiber_assignment_no_iip} but with a fiber assignment correction applied. All galaxies that are spectroscopically observed are weighted by the inverse probability of having been assigned a fiber.}
    \label{fig:fiber_assignment}
\end{figure*}

The bias in the selection of spectroscopic targets can be resolved by weighing each lens galaxy by its individual inverse probability (IIP) of being assigned a fiber, $\wsys \rightarrow \wsys \, w_{\mathrm{IIP}}$. The IIP can be estimated from Monte-Carlo simulations of the DESI fiber assignment pipeline on the photometric data and is one over the fraction of realizations in which each target is assigned a fiber. The idea of IIP-weighting is analogous to weighting by the pairwise-inverse-probability (PIP) weights for galaxy clustering \citep{Bianchi2018_MNRAS_481_2338}. The reason for using the individual instead of the pairwise probability is that in galaxy--galaxy lensing, only the lens is assigned a fiber, whereas for clustering both targets are subject to fiber incompleteness. In Fig.~\ref{fig:fiber_assignment}, we show that this IIP-weighting resolves any biases arising from fiber assignments: within the uncertainties the lensing signal around all targets and only the subset of targets with spectroscopic redshifts is identical to within $\lesssim 1 \%$ once the IIP-weights are applied. Finally, we also confirmed that this correction gives unbiased estimates even if the DESI spectroscopic coverage was very incomplete, i.e., only covering the sky with one pass. The results presented here are fully consistent with earlier studies by \cite{Makiya2022_JCAP_03_008} on the impact of fiber incompleteness on the galaxy--galaxy lensing amplitude around galaxies observed with the Prime Focus Spectrograph (PFS).

\subsection{Lens magnification}

\begin{figure*}
    \centering
    {\large\bf lens magnification bias}
    \includegraphics{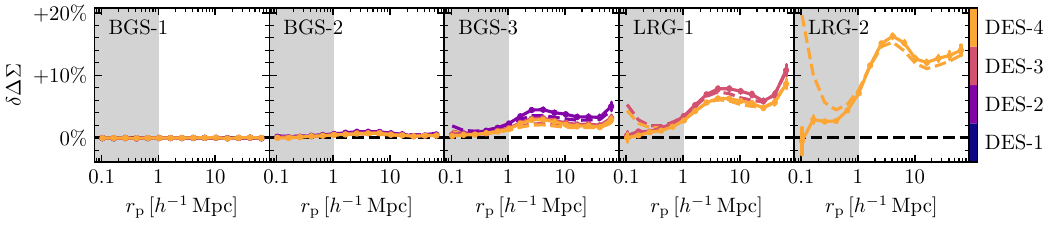}
    \includegraphics{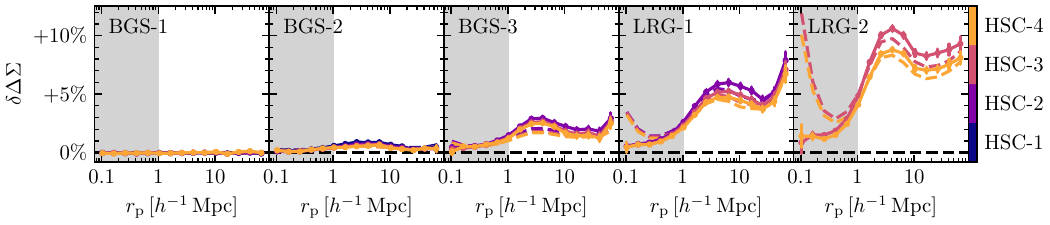}
    \includegraphics{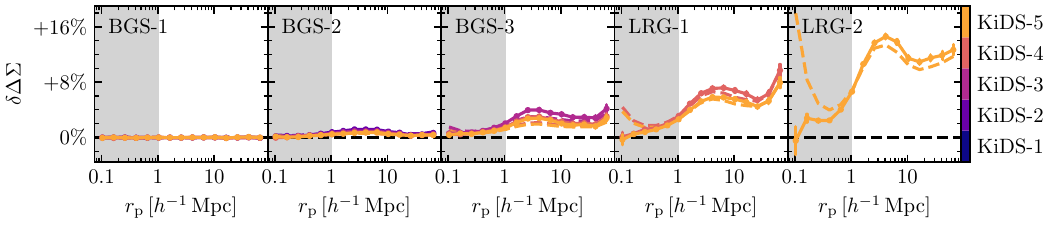}
    \caption{Relative impact of lens magnification on the measured galaxy--galaxy lensing amplitude. Dashed lines denote estimates of this contribution from eq.~\protect\eqref{eq:lens_magnification}.}
    \label{fig:lens_magnification}
\end{figure*}

\begin{figure*}
    \centering
    {\large\bf lens magnification bias}
    \includegraphics{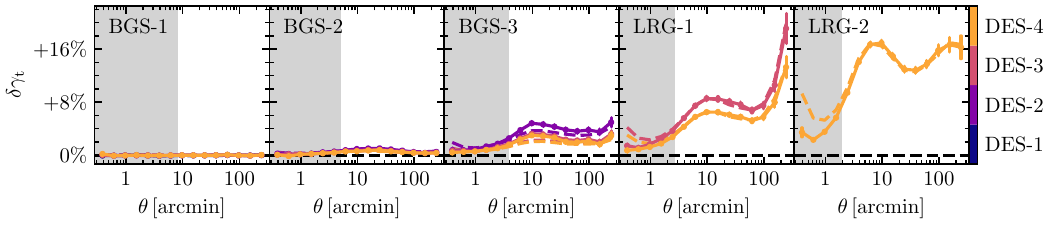}
    \caption{Similar to Fig.~\ref{fig:lens_magnification} but for $\gt$ instead of $\ds$.}
    \label{fig:lens_magnification_gt}
\end{figure*}

By default, our mock lensing catalogs contain the impact of magnification. As discussed previously, magnification affects the observed galaxy targets by a.) changing the apparent sizes and luminosities of targets and b.) affecting the apparent positions of galaxies. To undo effect a.), we run the target selection process for lenses and sources on the unlensed magnitudes. Similarly, one might be tempted to undo effect b.) by producing catalogs with angular coordinates unaffected by gravitational lensing. However, this approach does not work on small scales since the induced shear is a measure of the gravitational field at the apparent position of the source, not the intrinsic position in the absence of gravitational lensing. Thus, using intrinsic positions would lead to unreliable shear estimates on the scale of the typical deflection angle of sources, i.e. on $\mathrm{arcmin}$ scales. Instead, we recall that effect b.) changes the apparent number density of galaxies by a factor of $\mu^{-1}$. Thus, we can estimate the impact of removing effect b.) by multiplying weights $\wsys$ in the galaxy-galaxy lensing estimators, i.e. $w_{\mathrm{s}} \rightarrow w_{\mathrm{s}} \times \mu$ and $\wsys \rightarrow \wsys \times \mu$ for source and lens galaxies, respectively. Compared to using intrinsic coordinates, we checked that this method gives consistent results on larger scales and more realistic estimates of lens and source magnification on smaller scales.

To test the impact of lens magnification we compare galaxy--galaxy lensing measurements in mocks with and without magnification effects for DESI targets. As shown in Fig.~\ref{fig:lens_magnification} and \ref{fig:lens_magnification_gt}, lens magnification increases the measured lensing amplitude by up to $\sim 16\%$. The effect is strongest at intermediate to large scales, $\rp > 2 \mpch$ and increases sharply with lens redshift. We also find that the impact of lens magnification is not very sensitive to the choice of the lensing survey. Generally, galaxy-galaxy lensing with HSC has a smaller impact from lens magnification because it probes source galaxies at higher redshifts and therefore has a lower mean critical surface density $\sigmacrit$ for high-redshift lenses.

In the same figures, we also show the estimated impact of lens magnification using eq.~\eqref{eq:lens_magnification}. We estimate the impact of lens magnification by evaluating the above equation at the mean lens and source redshifts, i.e.
\begin{equation}
    \ds_{\mathrm{lm}} (\rp) = 2 (\alpha_{\mathrm{l}} - 1) \, \overline{\sigmacrit} \, \gamma_{\mathrm{LSS}} \left( \frac{\rp}{D_{\mathrm{C}} (\overline{\zl})}, \overline{\zl}, \overline{\zs} \right) \, ,
    \label{eq:lens_magnification}
\end{equation}
where the mean redshifts are taken over all lens-source pairs with the usual $\wls$ weights, i.e.
\begin{equation}
    \overline{z} = \frac{\sumls \wsys \wls z}{\sumls \wsys \wls}.
\end{equation}
Similarly, $\overline{\sigmacrit}$ is the mean critical surface density.

On small scales and large redshifts, there is a mismatch between the mocks and the predictions that is primarily due resolution issues below $1 \, \mpch$, as discussed in section \ref{subsec:gravitational_signal}. In all other cases, we observe a good agreement between the theoretical prediction and the impact of lens magnification measured in the mock catalogs. We observe a small $1 - 2\%$ offset for $\ds$ that is likely due to the approximation of assuming a single lens and source redshift. For $\gt$, predictions are in even better agreement with the mock results. We note that we assume perfect knowledge of the response of the lens field to magnification, i.e. we measured $\alpha_{\mathrm{l}}$ directly in the mock catalogs\footnote{We find $\alpha_{\mathrm{l}} = 0.91$, $1.58$, and $2.02$ for the three BGS samples as well as $\alpha_{\mathrm{l}} = 2.58$ and $2.26$ for the two LRG samples.}, as well as cosmology, particularly $\Omega_{\mathrm{m}}$ and $\sigma_8$. Studies focused on extracting cosmological parameters from galaxy--galaxy lensing will need to marginalize over uncertainties in these parameters.

\subsection{Source magnification}

\begin{figure*}
    \centering
    {\large\bf source magnification bias}
    \includegraphics{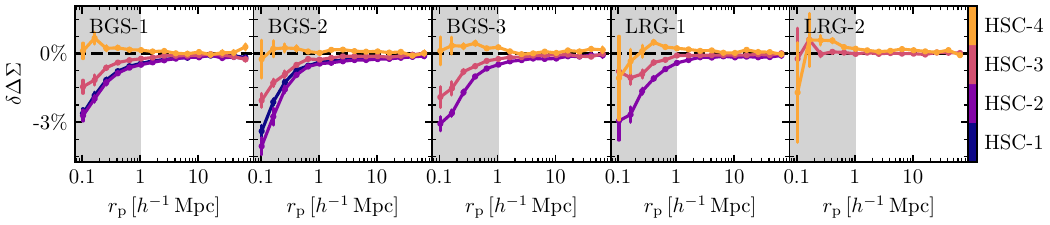}
    \caption{Relative impact of source magnification on the measured galaxy--galaxy lensing amplitude for HSC. Results for DES and KiDS are quantitatively similar.}
    \label{fig:source_magnification}
\end{figure*}

Source magnification describes the effect that sources behind lens galaxies are magnified by the gravitational potential associated with the lens galaxies themselves. Particularly, the effective number density of source galaxies is modulated with the lensing convergence map $\kappa (\mathbf{\theta}, \zs)$ in the same way that lens magnification affects the effective number density of lens galaxies. Similar to lens galaxies, the effective number of sources increases (decreases) with increasing $\kappa$ if $\alpha_{\mathrm{s}}$, defined for sources in analogy to $\alpha_{\mathrm{l}}$ for lenses in eq. (\ref{eq:alpha}), is larger (smaller) than unity. Our $\ds$ and $\gt$ estimators weigh each lens by the number of nearby sources. Implicitly, we assume these estimators are unbiased with respect to the mean $\ds$ and $\gt$ signal of all lenses. Source magnification violates this assumption because the number of sources is now modulated with $\kappa$ and by extension $\ds$ and $\gt$.

The impact of source magnification is challenging to estimate because it depends on bi-spectra \citep{Deshpande2020_AA_636_95} and because $\alpha_{\mathrm{s}}$ depends sensitively on a variety of lensing quality cuts. Fortunately, source magnification is generally a smaller effect than lens magnification, as shown in Fig.~\ref{fig:source_magnification}. For this figure, we compare mocks with and without magnification effects for source galaxies. Overall, the impact of source magnification can lead to an increase or decrease in the measured $\ds$ and $\gt$, depending on the effective value of $\alpha_{\mathrm{s}}$. However, at the scales most relevant for cosmological studies $\rp \gtrsim 5 \, \mpch$, we estimate the impact to be lower than $1\%$. We note that below those scales, our estimates using the Buzzard mocks might be biased due to the limited resolution of these simulations.

\subsection{Radially-dependent blending effects}

The average impact of blending on the measured shear has been estimated for each of the three surveys from image simulations \citep{Mandelbaum2018_MNRAS_481_3170, Giblin2021_AA_645_105, MacCrann2022_MNRAS_509_3371}. However, the mean corrections proposed by the different lensing surveys do not account for the systematic increase in source density near lens galaxies. Here, we estimate the impact of neglecting such radially dependent effects. As described in \cite{MacCrann2022_MNRAS_509_3371}, galaxy blends can have two effects on the measured shear. First, blending can bias the measured ellipticity of a population of galaxies irrespective of the ellipticities of the other galaxies involved in the blend. Second, the measured ellipticity of source galaxies can inherit some of the ellipticities of the galaxies they are blended with. The second case allows shear at one redshift to be ``transferred'' to the measured shear of galaxies at a second redshift. The latter effect leads to a shift in the effective source redshift distribution instead of just inducing a multiplicative shear bias $m$ \citep{MacCrann2022_MNRAS_509_3371}.

\begin{figure}
    \centering
    \includegraphics{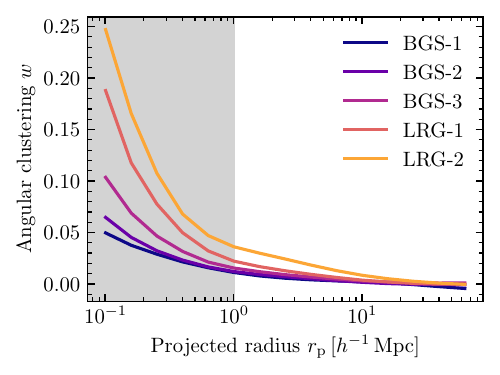}
    \caption{Angular clustering between DESI lens galaxy populations and DES source galaxies. Results for HSC and KiDS are qualitatively similar.}
    \label{fig:blending}
\end{figure}

Radially dependent blending effects might be introduced by the increased source density of galaxies physically associated with the lens. To first order, it should lead to a change in the blending induced multiplicative shear bias that should scale with the source number density, i.e. $\Delta m(\theta) \approx m_0 w(\theta)$ where $w(\theta)$ is the scale-dependent projected angular clustering of source galaxies, i.e., the average number of source galaxies around lenses compared to that around randoms, and $m_0$ the mean shear bias due to blending. In Fig.~\ref{fig:blending}, we show the angular clustering in DES sources around DESI lenses which is at most of order $w(\theta) \sim 5\%$ for $\rp > 1 \, \mpch$. Given that mean blending corrections are of the same order, $m_0 \sim \mathcal{O}(1 \%)$, this suggests that radially-dependent blending effects are negligible for galaxy--galaxy lensing measurements on those scales. However, excess blending may become relevant for studies focusing on very small scales, $\rp < 1 \, \mpch$.\footnote{In addition to affecting the measured shear of galaxies, blending may also affect the inferred magnitudes and thereby selection cuts and photometric redshifts \citep{Everett2022_ApJS_258_15}. We do not model these effects here.} At the same time, it is likely sub-dominant compared to the boost factor and the effect of the reduced shear approximation, as discussed next.

\subsection{Reduced shear approximation}

\begin{figure*}
    \centering
    {\large\bf reduced shear approximation}
    \includegraphics{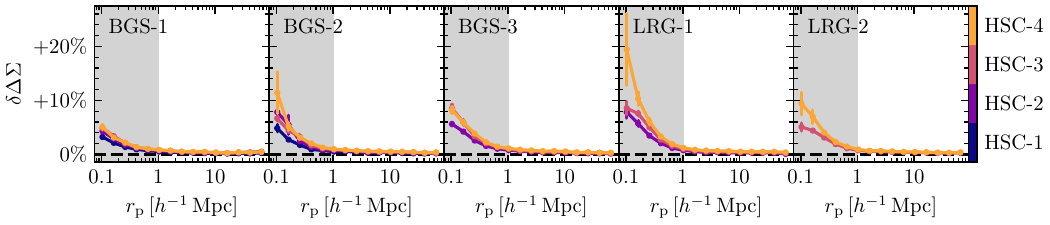}
    \caption{Relative impact of the reduced shear approximation on the interpretation of the galaxy--galaxy lensing amplitude for HSC. Assuming that the galaxy--galaxy lensing measurements are sensitive to the shear instead of the reduced shear leads to an overestimation of the intrinsic gravitational lensing signal, particularly on small scales. Results for DES and KiDS are quantitatively and quantitatively similar.}
    \label{fig:reduced_shear}
\end{figure*}

A common approximation in galaxy--galaxy lensing is to interpret the observable reduced shear $g$ as the not directly observable shear $\gamma$. In Fig.~\ref{fig:reduced_shear}, we show the impact of this assumption by comparing mocks where the measured ellipticity is derived from the shear $\gamma$ instead of the reduced shear $g$, as for the default mock catalogs. Results are shown for HSC but conclusions are similar for DES and KiDS, too. As expected, the tangential shear is higher if one uses the reduced shear since, on average, $(1 - \kappa)^{-1} > 1$. On the scales relevant for cosmological studies with DESI, i.e. $\rp > 1 \, \mpch$, the effect is very weak, of order $1 \%$ or less. The effect is expected to become stronger and potentially significant at current precision for smaller scales. However, the limited resolution of our simulations prevents us from drawing definitive conclusions.

\section{Discussion}
\label{sec:discussion}

In the previous section, we investigated several systematic effects in the galaxy--galaxy lensing estimators in detailed mock catalogs, specifically as they related to forthcoming galaxy--galaxy lensing measurements with DES, HSC, and KiDS source galaxies around DESI lens galaxies. For cosmological studies focusing on scales $r > 1 \, \mpch$, we find that IA, DESI fiber incompleteness, and lens magnification are the most relevant effects. We find that even if one only measures the lensing amplitude using sources sufficiently behind the lenses, the effects can be of order $20 \%$ and should be corrected or modeled. On smaller scales, $r < 1 \, \mpch$, boost factors and the reduced shear approximation may also be relevant. However, we caution the reader against drawing definitive conclusions given the limited resolution of our simulations. In contrast, the effect of source magnification, excess blending around DESI lenses as well as residual impacts of shear biases, and photometric redshift dilution are all likely not relevant given the expected precision of galaxy--galaxy lensing measurements with DESI. We note that for the latter two cases, shear biases and photometric redshift dilution, we assume that we have unbiased measurements of shear biases and perfectly characterized the photometric redshift distribution.

\begin{figure*}
    \centering
    \includegraphics{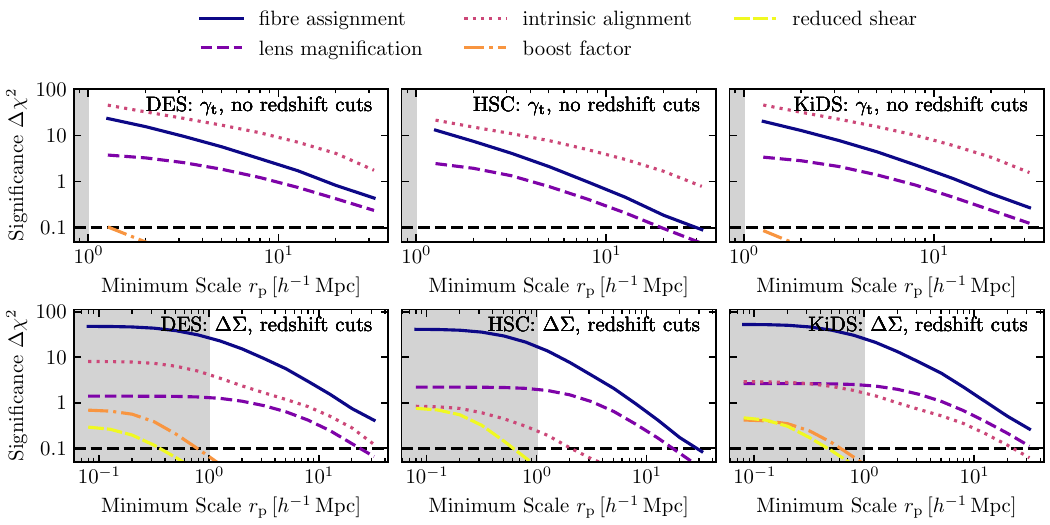}
    \caption{Estimates of the statistical significance in $\chi^2$ as a function of minimum-scale for the most important systematic effects in the galaxy--galaxy lensing measurements. We assume the uncertainties expected for the DESI year-1 galaxy--galaxy lensing analysis. Each column corresponds to one of the three lensing surveys. The upper row indicates sensitivities for measurements of $\gt$ and using all lens-source bin combinations whereas the lower row indicates measurements $\ds$ where only lens-source bin combinations are considered where sources are sufficiently behind lenses, as discussed in the text.}
    \label{fig:significance}
\end{figure*}

In the following, we estimate the statistical significance of these systematic effects. We quantify significance via the difference in $\chi^2$,
\begin{equation}
    \Delta\chi^2 = \Delta \gt \mathbf{C}_{\gt}^{-1} \Delta \gt^{\mathrm{T}} \,
\end{equation}
where $\Delta \gt$ is the absolute difference in $\gt$ induced by the systematic effect and $\mathbf{C}_{\gt}$ an estimate of the expected covariance matrix of galaxy--galaxy lensing measurements cross-correlation DESI year-1 data with weak lensing maps from DES Y3, HSC-Y1 and KiDS-1000 data \citep{Yuan2024_arXiv_2403_0915}. The definition for $\ds$ measurements is analogous. The above quantity gives us an upper limit on the biases one could realistically expect if such systematic effects are ignored. Assuming that biases are fully degenerate with changes in cosmological parameters, cosmological posterior estimates are biased by roughly $\sqrt{\Delta\chi^2} \sigma$.

For $\gt$, we consider all possible lens-source tomographic bin combinations, i.e., even those not included in the previous figures. Contrary, for $\ds$ we only consider bin combinations where lenses and sources are sufficiently far apart, i.e., $\zs > \zl + 0.2$, with $\zs$ and $\zs$ denoting the median lens and source redshifts, as defined previously. The motivation is that $\ds$ measurements will be used for a close to model-independent analysis of galaxy--galaxy lensing amplitudes with the different shear surveys, following \cite{Leauthaud2022_MNRAS_510_6150}. As shown below, such cuts can significantly mitigate the impact of IA which would otherwise have to be modeled very carefully.

In Fig.~\ref{fig:significance}, we show the expected significance of the most relevant systematic effects as a function of the minimum scale analyzed. We find that the effect of IA is most relevant if one measures $\gt$ for all lens-source bin combinations. This may include bin combinations where sources and lenses significantly overlap in redshift. In this case, the mean tangential shear may be dominated by IA. However, in the same figure, we also show that the impact of IA can be reduced dramatically if one employs suitable lens-source cuts, as done for the mock $\ds$ measurements. In this case, the impact of IA can be reduced to $\Delta\chi^2 \sim 1 - 3$, although it may still be significant. The residual impact is the strongest for DES owing to their relatively imprecise redshift estimates which leads to very wide tomographic bins. We recommend that all studies, particularly those not employing lens-source redshift cuts, should include a model for IA in their analysis. Finally, we note that our estimate of the IA contamination is based on the current observational constraints on IA via the Non-Linear Alignment model \citep[NLA;][]{Bridle2007_NJPh_9_444}. However, more complex models such as the \citep[TATT,][]{Blazek2019_PhRvD_100_3506} or halo-based models \citep{Fortuna2021_MNRAS_501_2983, VanAlfen2023_arXiv_2311_7374}, may describe the behavior of IA more accurately, especially on small-scale. Furthermore, the strength of IA is poorly constrained empirically. While our NLA amplitude, $A_{\textrm{IA}} = 0.5$, is consistent with current data, $A_{\textrm{IA}} = 1.0$ is not strongly excluded. Finally, it has been found in simulations that the IA amplitude increases with stellar mass and redshift \citep[see][and references therein]{Delgado2023_MNRAS_523_5899}. As a result, flux-limited lensing surveys may show stronger IA amplitudes at high-redshift \citep[see, e.g.,][]{DarkEnergySurveyandKiloDegreeSurveyCollaboration2023_OJAp_6_36}, leading to a further underestimation of the IA contamination, especially around LRGs. Ultimately, the impact of IA on galaxy--galaxy lensing may realistically be much higher than our mock results indicate. This further strengthens the conclusion that IA-contamination needs to be considered for galaxy--galaxy lensing estimates.

While we find that the impact of DESI fiber incompleteness is significant, even if one considers only regions that have completeness expected for the completed DESI survey, the effect can be fully mitigated by weighting the galaxies that have fibers by the individual inverse probability (IIP) to have been assigned a fiber. We recommend that all studies measuring galaxy--galaxy lensing incorporate this correction since the biases are otherwise very significant, as shown in Fig.~\ref{fig:significance}. Similarly, the effect of lens magnification is also statistically significant and cannot be mitigated via lens-source redshift cuts. Fortunately, the impact of lens magnification can be computed analytically and only requires input cosmological parameters and the sensitivity of DESI lenses to magnification. We suggest that all DESI galaxy--galaxy lensing studies employ either a correction for lens magnification assuming fixed cosmology or directly include cosmology-dependent lens magnification effects in their modeling pipeline.

Finally, on scales $\rp > 1 \, \mpch$, we find that the effects of boost factors and reduced shear are always subdominant, i.e., $\Delta\chi^2 < 0.1$, corresponding roughly to $<0.3 \sigma$. We note that our mock simulations suffer from resolution issues below $1 \, \mpch$. Additionally, we do not simulate effects like the impact of blending on galaxy colors or the relation between galaxy colors in photometric redshifts. This would likely be important on smaller scales where blending may be strong and there may exist strong color $\rp$-dependent gradients in the average color of nearby galaxies. Nevertheless, our results suggest that the reduced shear approximation and boost factors will become important for studies analyzing such smaller scales. Studies focusing on these scales should employ empirical or theoretical estimates of their impact.

\section{Conclusion}
\label{sec:conclusion}

Cross-correlating data from the DESI year-1 catalogs with weak lensing maps from leading imaging surveys will provide high signal-to-noise measurements of the galaxy--galaxy lensing effect around DESI galaxies. In turn, these measurements can be used to place tight constraints on the relation between galaxies and dark matter halos, fundamental cosmological parameters as well as tests of gravity on cosmological scales. To facilitate such studies, in this work, we estimate various systematic effects expected in the galaxy--galaxy lensing measurements with DESI using detailed ray-tracing simulations of gravitational lensing. Our main findings are as follows:

\begin{itemize}
    \item Without any corrections, the lensing amplitude measured around targets that receive DESI fibers is biased by up to $30 \%$ with respect to the expected amplitude if all members of a target class received fibers. Fortunately, this effect can be eliminated by weighting galaxies with fiber by the individual inverse probability (IIP) of receiving a fiber.
    \item The intrinsic alignment of source galaxies impacts galaxy--galaxy lensing measurements significantly. For lensing measurements with DES and KiDS, this effect is expected to be significant in DESI year-1 data even when employing conservative lens-source cuts.
    \item Lens magnification is important for galaxy--galaxy lensing measurements with DESI LRGs. Fortunately, the impact of lens magnification can be accurately predicted analytically once the sensitivity of DESI targets to magnification has been determined.
    \item At least for DESI year-1 data, boost factors and the reduced shear approximation do not bias galaxy--galaxy lensing measurements significantly on scales larger than $1 \, \mpch$. However, studies on smaller scales should evaluate the impact of these systematic effects.
    \item The impact of photometric redshift dilution and residual shear biases is negligible if one uses the estimators described in this work and those two effects have been accurately characterized. Similarly, the impact of source magnification and excess blending is likely negligible.
\end{itemize}

This work performs analysis choices for several upcoming studies cross-correlating DESI targets with weak lensing source catalogs. Among others, we plan to perform a unified analysis of cosmic shear, galaxy--galaxy lensing, and galaxy clustering, a so-called $3\times2$pt-analysis, with DESI, DES, HSC, and KiDS. Such an analysis is expected to result in leading constraints on cosmic structure growth and shed light on the so-called $S_8$-tension \citep{Abdalla2022_JHEAp_34_49}. Similarly, we will perform measurements of $\Delta\Sigma$ around common sets of DESI lenses. Since the measured $\Delta\Sigma$ amplitude has no dependence on the source catalog properties in the absence of the systematic effects discussed here, comparing measured $\Delta\Sigma$ amplitudes can be used as a powerful tool to empirically test the robustness of weak lensing measurements \citep[see][for an application to BOSS]{Leauthaud2022_MNRAS_510_6150, Amon2023_MNRAS_518_477}. The work presented here is a first step towards such future studies with DESI data.

\section*{Acknowledgments}

We thank Ji Yao, Alexandra Amon, Tomomi Sunayama, and the referee for insightful comments that improved this paper. We acknowledge the use of the lux supercomputer at UC Santa Cruz, funded by NSF MRI grant AST 1828315. We acknowledge support from the Leinweber Center for Theoretical Physics, NASA grant under contract 19-ATP19-0058, and DOE under contract DE-FG02-95ER40899.

This work made use of the following software packages: {\sc Astropy} \citep{AstropyCollaboration2013_AA_558_33}, {\sc bibmanager} \citep{Cubillos2020_zndo_25_7042}, {\sc dsigma} \citep{Lange2022_ascl_soft_4006}, {\sc NumPy} \citep{vanderWalt2011_CSE_13_22}, {\sc matplotlib} \citep{Hunter2007_CSE_9_90}, {\sc SciPy}, {\sc Spyder} and {\sc Setzer}.

This material is based upon work supported by the U.S. Department of Energy (DOE), Office of Science, Office of High-Energy Physics, under Contract No. DE–AC02–05CH11231, and by the National Energy Research Scientific Computing Center, a DOE Office of Science User Facility under the same contract. Additional support for DESI was provided by the U.S. National Science Foundation (NSF), Division of Astronomical Sciences under Contract No. AST-0950945 to the NSF’s National Optical-Infrared Astronomy Research Laboratory; the Science and Technology Facilities Council of the United Kingdom; the Gordon and Betty Moore Foundation; the Heising-Simons Foundation; the French Alternative Energies and Atomic Energy Commission (CEA); the National Council of Science and Technology of Mexico (CONACYT); the Ministry of Science and Innovation of Spain (MICINN), and by the DESI Member Institutions: \url{https://www.desi.lbl.gov/collaborating-institutions}. Any opinions, findings, and conclusions or recommendations expressed in this material are those of the author(s) and do not necessarily reflect the views of the U. S. National Science Foundation, the U. S. Department of Energy, or any of the listed funding agencies.

The authors are honored to be permitted to conduct scientific research on Iolkam Du’ag (Kitt Peak), a mountain with particular significance to the Tohono O’odham Nation.

\section*{Data Availability}

All data points shown in the published graph are available at \url{https://zenodo.org/doi/10.5281/zenodo.10934082}. Additionally, all the code used to generate plots in this work will be made available on GitHub upon publication. All data products produced in this work are available upon reasonable request to the lead author.

\bibliographystyle{mnras}
\bibliography{bibliography}

\section*{Affiliations}
\noindent
{\footnotesize $^{1}$ Department of Physics, University of Michigan, Ann Arbor, MI 48109, USA\\
$^{2}$ Centre for Astrophysics \& Supercomputing, Swinburne University of Technology, P.O. Box 218, Hawthorn, VIC 3122, Australia\\
$^{3}$ Max Planck Institute for Extraterrestrial Physics, Gie\ss enbachstra\ss e 1, 85748 Garching, Germany\\
$^{4}$ Universit\"ats-Sternwarte M\"unchen, Scheinerstra\ss e 1, 81679 Munich, Germany\\
$^{5}$ Department of Physics \& Astronomy, University College London, Gower Street, London, WC1E 6BT, UK\\
$^{6}$ Lawrence Berkeley National Laboratory, 1 Cyclotron Road, Berkeley, CA 94720, USA\\
$^{7}$ Argonne National Laboratory, High-Energy Physics Division, 9700 S. Cass Avenue, Argonne, IL 60439, USA\\
$^{8}$ Department of Physics, The University of Texas at Dallas, Richardson, TX 75080, USA\\
$^{9}$ University of California, Berkeley, 110 Sproul Hall \#5800 Berkeley, CA 94720, USA\\
$^{10}$ Department of Astronomy and Astrophysics, UCO/Lick Observatory, University of California, 1156 High Street, Santa Cruz, CA 95064, USA\\
$^{11}$ Institute of Cosmology and Gravitation, University of Portsmouth, Dennis Sciama Building, Portsmouth, PO1 3FX, UK\\
$^{12}$ Aix Marseille Univ, CNRS, CNES, LAM, Marseille, France\\
$^{13}$ Department of Physics and Astronomy, University of Waterloo, 200 University Ave W, Waterloo, ON N2L 3G1, Canada\\
$^{14}$ Perimeter Institute for Theoretical Physics, 31 Caroline St. North, Waterloo, ON N2L 2Y5, Canada\\
$^{15}$ Waterloo Centre for Astrophysics, University of Waterloo, 200 University Ave W, Waterloo, ON N2L 3G1, Canada\\
$^{16}$ Department of Astronomy and Astrophysics, University of California, Santa Cruz, 1156 High Street, Santa Cruz, CA 95065, USA\\
$^{17}$ Institute for Astronomy, University of Edinburgh, Royal Observatory, Blackford Hill, Edinburgh EH9 3HJ, UK\\
$^{18}$ Ruhr University Bochum, Faculty of Physics and Astronomy, Astronomical Institute (AIRUB), German Centre for Cosmological Lensing, 44780 Bochum, Germany\\
$^{19}$ The Ohio State University, Columbus, 43210 OH, USA\\
$^{20}$ Department of Physics and Astronomy, Sejong University, Seoul, 143-747, Korea\\
$^{21}$ School of Mathematics and Physics, University of Queensland, 4072, Australia\\
$^{22}$ SLAC National Accelerator Laboratory, Menlo Park, CA 94305, USA\\
$^{23}$ Physics Dept., Boston University, 590 Commonwealth Avenue, Boston, MA 02215, USA\\
$^{24}$ Instituto de F\'{\i}sica, Universidad Nacional Aut\'{o}noma de M\'{e}xico,  Cd. de M\'{e}xico  C.P. 04510,  M\'{e}xico\\
$^{25}$ Kavli Institute for Particle Astrophysics and Cosmology, Stanford University, Menlo Park, CA 94305, USA\\
$^{26}$ Institut de F\'{i}sica d’Altes Energies (IFAE), The Barcelona Institute of Science and Technology, Campus UAB, 08193 Bellaterra Barcelona, Spain\\
$^{27}$ Departamento de F\'isica, Universidad de los Andes, Cra. 1 No. 18A-10, Edificio Ip, CP 111711, Bogot\'a, Colombia\\
$^{28}$ Observatorio Astron\'omico, Universidad de los Andes, Cra. 1 No. 18A-10, Edificio H, CP 111711 Bogot\'a, Colombia\\
$^{29}$ Institut d'Estudis Espacials de Catalunya (IEEC), 08034 Barcelona, Spain\\
$^{30}$ Institute of Space Sciences, ICE-CSIC, Campus UAB, Carrer de Can Magrans s/n, 08913 Bellaterra, Barcelona, Spain\\
$^{31}$ NSF NOIRLab, 950 N. Cherry Ave., Tucson, AZ 85719, USA\\
$^{32}$ Department of Physics, Southern Methodist University, 3215 Daniel Avenue, Dallas, TX 75275, USA\\
$^{33}$ Departament de F\'{i}sica, Serra H\'{u}nter, Universitat Aut\`{o}noma de Barcelona, 08193 Bellaterra (Barcelona), Spain\\
$^{34}$ Instituci\'{o} Catalana de Recerca i Estudis Avan\c{c}ats, Passeig de Llu\'{\i}s Companys, 23, 08010 Barcelona, Spain\\
$^{35}$ Department of Physics and Astronomy, Siena College, 515 Loudon Road, Loudonville, NY 12211, USA\\
$^{36}$ Department of Physics and Astronomy, University of Sussex, Brighton BN1 9QH, U.K\\
$^{37}$ Department of Physics \& Astronomy, University  of Wyoming, 1000 E. University, Dept.~3905, Laramie, WY 82071, USA\\
$^{38}$ National Astronomical Observatories, Chinese Academy of Sciences, A20 Datun Rd., Chaoyang District, Beijing, 100012, P.R. China\\
$^{39}$ Departamento de F\'{i}sica, Universidad de Guanajuato - DCI, C.P. 37150, Leon, Guanajuato, M\'{e}xico\\
$^{40}$ Instituto Avanzado de Cosmolog\'{\i}a A.~C., San Marcos 11 - Atenas 202. Magdalena Contreras, 10720. Ciudad de M\'{e}xico, M\'{e}xico\\
$^{41}$ IRFU, CEA, Universit\'{e} Paris-Saclay, F-91191 Gif-sur-Yvette, France\\
$^{42}$ Space Sciences Laboratory, University of California, Berkeley, 7 Gauss Way, Berkeley, CA  94720, USA\\
$^{43}$ Department of Physics, Kansas State University, 116 Cardwell Hall, Manhattan, KS 66506, USA\\
$^{44}$ CIEMAT, Avenida Complutense 40, E-28040 Madrid, Spain\\
$^{45}$ University of Michigan, Ann Arbor, MI 48109, USA\\
$^{46}$ Department of Physics \& Astronomy, Ohio University, Athens, OH 45701, USA\\
$^{47}$ Physics Department, Stanford University, Stanford, CA 93405, USA\\
}

\end{document}